\newif\ifjournal
\journalfalse 
\newif\ifUSEMULTIBIB
\USEMULTIBIBfalse

\ifjournal
	\documentclass[
	 			preprint,     		superscriptaddress, 																longbibliography,
		amsmath, amssymb,
		aps,  prb,   		floatfix,
		linenumbers     
	]{revtex4-1} 
\else
	\documentclass[
		reprint,
		superscriptaddress, 
		longbibliography,
		amsmath, 
		amssymb,
		aps,  
		prb, 
		floatfix,
	]{revtex4-1} 
\fi

\usepackage[utf8]{inputenc}   \usepackage[T1]{fontenc}
\usepackage{textcomp}             
\usepackage{bm}					\usepackage{enumerate}
\usepackage[hidelinks]{hyperref}    \usepackage[mathlines]{lineno}	  \pdfoutput=1 					       
\usepackage{graphicx}			
\usepackage{multirow}
\usepackage{dcolumn}			\usepackage[usenames, dvipsnames]{colortbl} 
\usepackage[normalem]{ulem}

\ifUSEMULTIBIB
	\usepackage{multibib}
	\newcites{Methods}{Methods}
								\else 	\newcommand{\citeMethods}{\cite}
	\newcommand{\onlineciteMethods}{\onlinecite}
	
	\newcommand{\nociteMethods}{\nocite}
	\newcommand{\bibliographystyleMethods}{\bibliographystyle}
	\newcommand{\bibliographyMethods}{\bibliography}
\fi

\newcommand{\ket}[1]{\left|#1\right>}

\newcommand{\B}{\ket{\mathrm{B}}}

\begin{document}

\makeatletter
\renewcommand{\fnum@figure}{\textbf{Figure~\thefigure}}
\makeatother

\title{To catch and reverse a quantum jump mid-flight}

\author{Z.K. Minev}
 \affiliation{Department of Applied Physics, Yale University, New Haven, Connecticut 06511, USA}
  \author{S.O. Mundhada}
   \affiliation{Department of Applied Physics, Yale University, New Haven, Connecticut 06511, USA}
\author{S. Shankar}
   \affiliation{Department of Applied Physics, Yale University, New Haven, Connecticut 06511, USA}
\author{P. Reinhold}
   \affiliation{Department of Applied Physics, Yale University, New Haven, Connecticut 06511, USA}
\author{R. Guti\'errez-J\'auregui}
   \affiliation{The Dodd-Walls Centre for Photonic and Quantum Technologies, Department of Physics, University of Auckland, Private Bag 92019, Auckland, New Zealand}
\author{R.J. Schoelkopf}
   \affiliation{Department of Applied Physics, Yale University, New Haven, Connecticut 06511, USA}
\author{M. Mirrahimi}
	\affiliation{Yale Quantum Institute, Yale University, New Haven, Connecticut 06520, USA}
	\affiliation{QUANTIC team, INRIA de Paris, 2 Rue Simone Iff, 75012 Paris, France}
\author{H.J. Carmichael}
   \affiliation{The Dodd-Walls Centre for Photonic and Quantum Technologies, Department of Physics, University of Auckland, Private Bag 92019, Auckland, New Zealand}
\author{M.H. Devoret}
   \affiliation{Department of Applied Physics, Yale University, New Haven, Connecticut 06511, USA}

\date{\today}

\maketitle

\onecolumngrid
\textbf{ 
Quantum physics was invented to account for  
 two fundamental features of measurement results — their discreetness and  randomness. Emblematic of these features is Bohr’s idea of quantum jumps between two discrete energy levels of an atom\cite{Bohr1913}. Experimentally, quantum jumps were first observed in an  atomic ion driven by a weak deterministic force while under strong continuous energy measurement\cite{Nagourney1986, Sauter1986, Bergquist1986}. The times at which the discontinuous jump transitions occur are reputed to be fundamentally unpredictable. 
Can there be, despite the indeterminism of quantum physics, a possibility to know if a quantum jump is about to occur or not?
Here, we answer this question affirmatively by experimentally demonstrating that the  jump from the ground to an excited state of a superconducting artificial three-level atom can be tracked as it follows a predictable ``flight,'' 
by monitoring the population of an auxiliary energy level coupled to the ground state.  The experimental results demonstrate that the jump evolution when completed is continuous, coherent, and deterministic. Furthermore, exploiting these features and using real-time monitoring and feedback, we catch and reverse a quantum jump mid-flight, thus deterministically preventing its completion. 
Our results, which agree with theoretical predictions essentially without adjustable parameters,  support the modern quantum trajectory theory\cite{Carmichael1993, Gardiner1992-original-traj, Dalibard1992-original-traj, Plenio1998, Korotkov1999-original-traj} and  provide new ground for the exploration of real-time intervention techniques in the control of quantum systems, such as early detection of error syndromes.
}
\vspace{1em}
\twocolumngrid

Bohr conceived of quantum jumps \cite{Bohr1913} in 1913, and while Einstein elevated their hypothesis to the level of a quantitative rule with his AB coefficient theory\cite{Einstein1916, Einstein1917}, Schr\"{o}dinger strongly objected to their existence\cite{Schrodinger1952}. The nature and existence of quantum jumps remained a subject of controversy for seven decades until they were directly observed in a single system \cite{Nagourney1986, Sauter1986, Bergquist1986}. 
Since then, quantum jumps have been observed in a variety of  atomic\cite{Basche1995, Peil1999, Gleyzes2007, Guerlin2007} and solid-state \cite{Jelezko2002, Neumann2010, Robledo2011, Vijay2011, Hatridge2013} systems.  
Recently,  quantum jumps have been recognized as an essential phenomenon in quantum feedback  control \cite{Deleglise2008, Sayrin2001}, and in particular, for detecting and correcting decoherence-induced errors in quantum information systems \cite{Sun2013, Ofek2016}.

\begin{figure}[!ht]
\label{fig:setup}
\begin{centering}
\ifjournal
	\includegraphics[width=89mm]{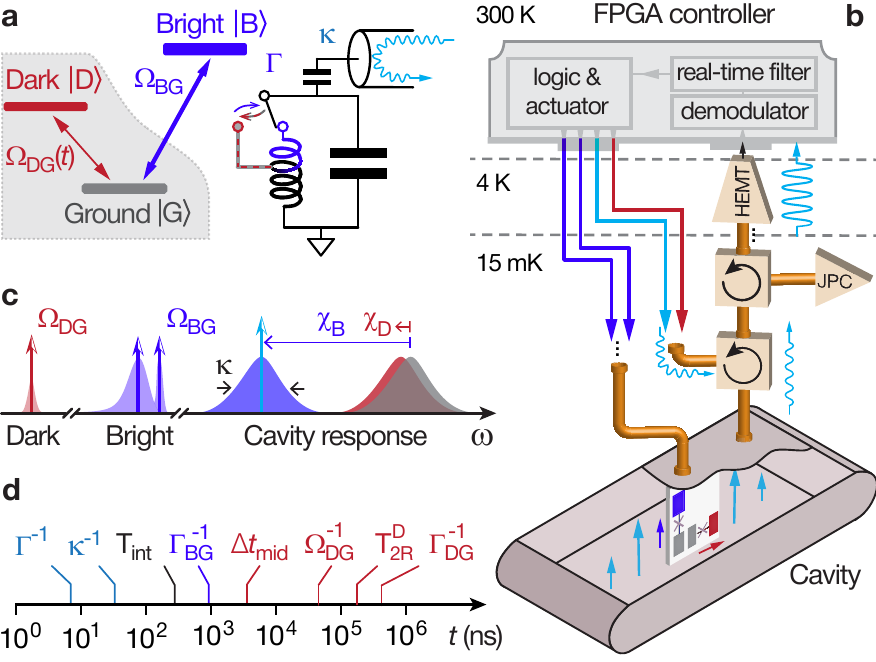}
\else
	\includegraphics[width=89mm]{1_setup.pdf}
\fi
\caption{
\textbf{Principle of the experiment.} 
\textbf{a,} 
Three-level atom possessing a hidden transition (shaded region) between its ground $\ket{\mathrm{G}}$ and dark $\ket{\mathrm{D}}$ state, driven by Rabi drive $\Omega_\mathrm{DG}(t)$.
Quantum jumps between $\ket{\mathrm{G}}$ and $\ket{\mathrm{D}}$ are indirectly monitored by a stronger Rabi drive $\Omega_\mathrm{BG}$ between $\ket{\mathrm{G}}$ and the bright state $\ket{\mathrm{B}}$, whose occupancy is continuously monitored at rate $\Gamma$ by an auxiliary oscillator (LC circuit on right), itself measured in reflection by  continuous-wave microwave light (depicted in light blue). 
When the atom is in $\ket{\mathrm{B}}$,  the LC circuit resonance frequency shifts to a lower frequency than when the atom is in $\ket{\mathrm{G}}$ or $\ket{\mathrm{D}}$ (effect schematically represented by switch). 
Therefore, the probe tone performs a $\ket{\mathrm{B}}$/not-$\ket{\mathrm{B}}$ measurement on the atom, and is blind to any superposition of  $\ket{\mathrm{G}}$ and $\ket{\mathrm{D}}$.
\textbf{b,}  
The actual atom and LC oscillator used in the experiment is a superconducting circuit consisting of two strongly-hybridized transmon qubits placed inside a readout resonator cavity at 15\,mK.
Control signals for the atom and cavity are supplied by a room-temperature field-programmable gate array (FPGA) controller.
This fast electronics  monitors the reflected signal from the cavity, and after demodulation and filtering, actuates the control signals.
The amplifier chain includes circulators (curved arrows) and amplifiers (triangles and trapezoids).
\textbf{c,}  
Frequency landscape of atom and cavity responses, overlaid with the control tones shown as vertical arrows.
The cavity pull $\chi$ of the atom is nearly identical for $\ket{\mathrm{G}}$ and $\ket{\mathrm{D}}$, but  markedly distinct for $\ket{\mathrm{B}}$.
The BG drive is bi-chromatic in order to address the bright transition independently of the cavity state. \textbf{d,} Hierarchy of timescales involved in the experiment, which are required to span 5 orders of magnitude. Symbols explained in text,  and  summarized in Extended Data Table~\ref{tab:summary-timescales}.
}
\end{centering}
\end{figure}

Here, we focus on the canonical  case of quantum jumps between two levels indirectly monitored by a third --- the case that corresponds to the original observation of quantum jumps in atomic physics\cite{Nagourney1986, Sauter1986, Bergquist1986}, see the level diagram of Fig.~1a.  
A surprising prediction emerges according to quantum trajectory theory:\cite{Carmichael1993, Porrati1987, Ruskov2007} not only does the state of the system evolve continuously during the jump between the ground $\ket{\mathrm{G}}$ and excited $\ket{\mathrm{D}}$ state, but  it is predicted that there is always a latency period prior to the jump, during which it is possible to acquire a signal that warns of the imminent occurrence of the jump (see Supplement Sec.~IIA).
This advance warning signal consists of a rare, particular lull in the excitation of the ancilla state $\ket{\mathrm{B}}$.
The acquisition of this signal requires the time-resolved, fully efficient detection of  \textit{every} de-excitation of $\ket{\mathrm{B}}$.
Exploiting the specific advantages of superconducting artificial atoms and their  quantum-limited readout chain, we designed an experiment that implements with maximum fidelity and minimum delay the detection of the  advance warning signal occurring before the quantum jump  (see rest of Fig.~1).

First, we developed a superconducting artificial atom with the necessary V-shape level structure (see Fig.~1a and Methods). It consists, besides the ground level $\ket{\mathrm G}$,  of one protected, dark level $\ket{\mathrm D}$ ---  engineered to not couple to any dissipative environment or any measurement apparatus 
 ---  and one ancilla level $\ket{\mathrm B}$, whose occupation is monitored at rate $\Gamma$.
Quantum jumps between $\left|\mathrm G\right>$ and $\left|\mathrm D\right>$ are induced by a weak Rabi drive $\Omega_{\mathrm{DG}}$ --- although this drive might eventually be turned off during the jump, as explained later.
Since a direct measurement of the dark level is not feasible, the jumps are monitored using the Dehmelt shelving scheme\cite{Nagourney1986}.
Thus, the occupation of $\left|\mathrm G\right>$  is linked to  that of $\left|\mathrm B\right>$ by the strong Rabi drive $\Omega_{\mathrm{BG}}$ ($ \Omega_{\mathrm{DG}} \ll \Omega_{\mathrm{BG}} \ll \Gamma $). 
In the atomic physics shelving scheme\cite{Nagourney1986, Sauter1986, Bergquist1986}, an excitation to $\ket{\mathrm{B}}$ is recorded by detecting the emitted photons from $\ket{\mathrm{B}}$ with a photodetector. From the detection events --- referred to in the following as ``clicks'' --- one infers the occupation of  $\ket{\mathrm{G}}$. On the other hand, from the prolonged absence of clicks (to be defined precisely below; see also Supplement Sec.~II), one infers that a quantum jump from $\ket{\mathrm{G}}$ to $\ket{\mathrm{D}}$ has occurred.
Due to the poor collection efficiency and dead-time of photon counters in atomic physics\cite{Volz2011}, it is exceedingly difficult to detect every individual click required to faithfully register the origin in time of the advance warning signal.
However, superconducting systems present the advantage of high collection efficiencies\cite{Riste2013, Murch2013, Weber2014}, as  their microwave photons are emitted into one-dimensional waveguides  and are detected with the same detection efficiencies as optical photons.
Furthermore, rather than monitoring the direct fluorescence of the $\ket{\mathrm{B}}$ state, we monitor its occupation by dispersively coupling it to an ancilla readout cavity. 
This further improves the fidelity of the detection of the  de-excitation from $\ket{\mathrm{B}}$ (effective collection  efficiency of  photons emitted from $\ket{\mathrm{B}}$).

\begin{figure*}
\label{fig:jumps}
\begin{centering}
\ifjournal
	\includegraphics[width=147mm]{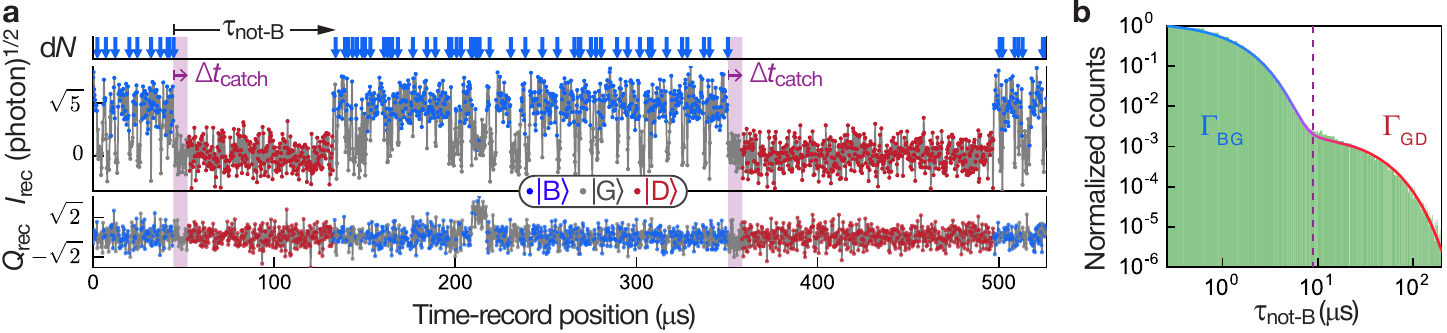}
\else
	\includegraphics[width=147mm]{2_jumps.pdf}
\fi
\caption{
\textbf{Unconditioned monitoring of quantum jumps in the 3-level system.} 
\textbf{a,} 
Typical measurement of integrated, with duration $T_\mathrm{int},$ quadratures $I_\mathrm{rec}$ and $Q_\mathrm{rec}$ of signal reflected from readout cavity  as a function of time. 
The color of the dots (see legend) denotes the state of the atom estimated by a real-time filter implemented with the FPGAs (Methods). 
On top, the vertical arrows indicate ``click'' events ($\mathrm{d}N$) corresponding to the inferred state changing from $\ket{\mathrm{B}}$ to not-$\ket{\mathrm{B}}$.
The symbol $\tau_{\operatorname{not-B}}$ corresponds to the time spent in not-$\ket{\mathrm{B}}$, which is the time between two clicks minus the last duration spent in $\ket{\mathrm{B}}$.  
An advance warning that a jump to $\ket{\mathrm{D}}$ is occurring is triggered when \textit{no} click has been observed for a duration $\Delta t_\mathrm{catch}$, which is chosen between 1 and 12$\,\mathrm{\mu s}$ at the start of the experiment.
\textbf{b,}  Log-log plot of the histogram of $\tau_{\operatorname{not-B}}$ (shaded green) for 3.2 s of continuous data of the type of panel (a).
Solid line is a bi-exponential fit defining jump rates $\Gamma_{\mathrm {BG}}= \left( 0.99\pm0.06\,\mathrm{\mu s}\right)^{-1}$ and  $\Gamma_{\mathrm{GD}}= \left(30.8\pm0.4\,\mathrm{\mu s}\right)^{-1}$.  
}
\end{centering}
\end{figure*}

The readout cavity, schematically depicted in Fig.~1a by an LC circuit, is resonant at $\omega_\mathrm{C} = 8979.64\,\mathrm{MHz}$ and cooled to 15 mK. 
Its dispersive coupling to the atom results in a conditional shift of its resonance frequency by $\chi_\mathrm{B}/2\pi = -5.08 \pm 0.2\,\mathrm{MHz}$ ($\chi_\mathrm{D}/2\pi = -0.33 \pm 0.08\,\mathrm{MHz}$) when the atom is in $\ket{\mathrm{B}}$ ($\ket{\mathrm{D}}$), see Fig.~1c. 
The engineered large asymmetry between $\chi_\mathrm{B}$ and $\chi_\mathrm{D}$ together with the cavity coupling rate to the output waveguide, $\kappa/2\pi = 3.62 \pm 0.05\,\mathrm{MHz}$, renders the cavity response markedly resolving for $\ket{\mathrm{B}}$ vs.~not-$\ket{\mathrm{B}}$, yet non-resolving\cite{Riste2013}  for $\ket{\mathrm{G}}$ vs.~$\ket{\mathrm{D}}$, thus preventing information about the dark transition from reaching the environment.   When probing the cavity response at $\omega_\mathrm{C} - \chi_\mathrm{B}$, the cavity  either remains empty, when the atom is in $\ket{\mathrm{G}}$ or $\ket{\mathrm{D}}$, or fills with $\bar{n}=5 \pm 0.2$ photons  when the atom is in $\ket{\mathrm{B}}$.
This readout scheme yields a transduction of the $\ket{\mathrm{B}}$-occupancy signal with five-fold amplification, which is an important advantage to overcome the noise of the following amplification stages. To summarize, in this readout scheme, the  cavity probe inquires: Is the atom in $\ket{\mathrm{B}}$ or not? The time needed to arrive at an answer with a confidence level of 68\% (signal-to-noise ratio of 1) is $\Gamma^{-1} \approx 1/\left(\kappa \bar{n}\right) = 8.8~\mathrm{ns}$ for an ideal amplifier chain (Supplement Sec.~IIIC).

Importantly, the engineered near-zero coupling between the cavity and the $\ket{\mathrm{D}}$ state protects the $\ket{\mathrm{D}}$ state from harmful effects, including Purcell relaxation, photon shot-noise dephasing, and the yet essentially unexplained residual measurement-induced relaxation in superconducting qubits (Supplement Sec.~I).We have measured the following coherence times for the $\ket{\mathrm{D}}$ state: energy relaxation $T_1^\mathrm{D} = 116 \pm 5~\mathrm{\mu s}$, Ramsey coherence $T_{\mathrm{2R}}^\mathrm{D} = 120 \pm 5~\mathrm{\mu s}$, and Hahn echo $T_\mathrm{2E}^\mathrm{D} = 162 \pm 6~\mathrm{\mu s}$. 
While protected, the $\ket{\mathrm{D}}$ state is indirectly quantum-non-demolition (QND) read out by the combination of the V-structure, the drive between $\ket{\mathrm{G}}$ and $\ket{\mathrm{B}}$, and the fast $\ket{\mathrm{B}}$-state monitoring. In practice, we can access the population of $\ket{\mathrm{D}}$ using an 80\,ns unitary rotation followed by a projective measurement of $\ket{\mathrm{B}}$ (Methods). 
 
Once the state of the readout cavity is imprinted with information about the occupation of $\ket{\mathrm{B}}$, photons leak through the cavity output port into a superconducting waveguide, which is connected to the amplification chain, see Fig.~1b, where they are amplified by a factor of $10^{12}$. 
The first stage of amplification is a quantum-limited Josephson parametric converter (JPC), which is followed by a high-electron-mobility transistor (HEMT) amplifier at 4 K.
The overall efficiency of the amplification chain is $\eta = 0.33 \pm 0.03$, which includes all possible loss of information, such as due to photon loss, thermal photons, jitter, etc. (see Methods). At room temperature, the heterodyne signal is demodulated by a home-built field-programmable gate array (FPGA) controller, with a 4\,ns clock period for logic operations. 
The measurement record  consists of a time series of two quadrature outcomes, $I_\mathrm{rec}$ and $Q_\mathrm{rec}$,  every 260 ns,
which is the integration time $T_\mathrm{int}$,
from which the FPGA controller estimates the state of the atom in real time.
To reduce the influence of noise, the controller applies a real-time, hysteretic IQ filter (see Methods), and then, from the estimated atom state,  the control drives of the atom and readout cavity are actuated, realizing feedback control.

Having described  the setup of the experiment, we proceed to report its  results. 
The field reflected out of the cavity is monitored in a free-running protocol, for which the atom is subject to the continuous Rabi drives $\Omega_\mathrm{BG}$ and $\Omega_\mathrm{DG}$, as depicted in Fig.~1.
Figure~2a shows a typical trace of the measurement record, displaying the quantum jumps of our three-level artificial atom. 
For most of the displayed duration of the record, $I_\mathrm{rec}$  switches rapidly between a low and high value, corresponding to approximately 0 ($\ket{\mathrm{G}}$ or $\ket{\mathrm{D}}$) and 5 ($\ket{\mathrm{B}}$) photons in the cavity, respectively. Spikes in $Q_\mathrm{rec}$, such as the one at $t=210\,\mathrm{\mu s}$ are recognized by the FPGA logic as a short-lived excursion of the atom to a higher excited state (Methods).
The corresponding  state of the atom, estimated by the FPGA controller, is depicted by the  color of the dots.
A change from $\ket{\mathrm{B}}$  to not-$\ket{\mathrm{B}}$ is equivalent to a ``click'' event, in that it corresponds to the emission of a photon from $\ket{\mathrm{B}}$ to $\ket{\mathrm{G}}$, whose occurrence time is shown by the vertical  arrows in the inferred record $\mathrm dN\left(t\right)$ (top). 
We could also indicate upward transitions from $\ket{\mathrm{G}}$ to $\ket{\mathrm{B}}$, corresponding to photon absorption events (not emphasized here), which would not be detectable in the atomic case.

In the example record, the detection of clicks stops completely at  $t=45\,\mathrm{\mu s}$, which reveals a quantum jump from $\ket{\mathrm{G}}$  to $\ket{\mathrm{D}}$, see Methods for operational definition of quantum jumps. 
The state $\ket{\mathrm{D}}$ survives for $90\, \mathrm{\mu s}$ before the atom returns to $\ket{\mathrm{G}}$ at $t =  135\, \mathrm{\mu s}$, when  the rapid switching between $\ket{\mathrm{G}}$ and $\ket{\mathrm{B}}$  resumes until a second quantum jump to the dark state occurs at $t = 350\, \mathrm{\mu s}$.
Thus, the record presents jumps from $\ket{\mathrm{G}}$ to $\ket{\mathrm{D}}$ in the form of click interruptions. These ``outer'' jumps occur on a much longer time scale than the ``inner'' jumps from $\ket{\mathrm{G}}$ to $\ket{\mathrm{B}}$.

In Fig.~2b, which is based on the continuous tracking of the quantum jumps for 3.2\,s, a histogram of the time spent in not-$\ket{\mathrm{B}}$, $\tau_{\operatorname{not-B}}$, is shown (See Extended Data Fig.~\ref{fig:B-wait-time} for the time spent in $\B$). 
The panel further shows a fit of the histogram by a bi-exponential curve that models two interleaved Poisson processes. 
This yields the average time the atom rests in $\ket{\mathrm{G}}$ before an excitation to $\ket{\mathrm{B}}$, $\Gamma_{\mathrm{BG}}^{-1} = 0.99\pm0.06\, \mathrm{\mu s}$, and the average time the atom stays up in $\ket{\mathrm{D}}$ before returning to $\ket{\mathrm{G}}$ and being detected, $\Gamma_{\mathrm{GD}}^{-1} = 30.8\pm0.4\, \mathrm{\mu s}$.
The average time between two consecutive $\ket{\mathrm{G}}$ to $\ket{\mathrm{D}}$ jumps  is $\Gamma_{\mathrm{DG}}^{-1} = 220\pm 5\, \mathrm{\mu s}$. The corresponding rates depend on the atom drive amplitudes ($\Omega_\mathrm{DG}$ and $\Omega_\mathrm{BG}$) and the measurement rate $\Gamma$ (Supplement Sec.~II).
Crucially, all the rates in the system must be distributed over a minimum of 5 orders of magnitude, as shown in Fig.~1d.

\begin{figure}
\begin{centering}
\ifjournal
	\includegraphics[width=89mm]{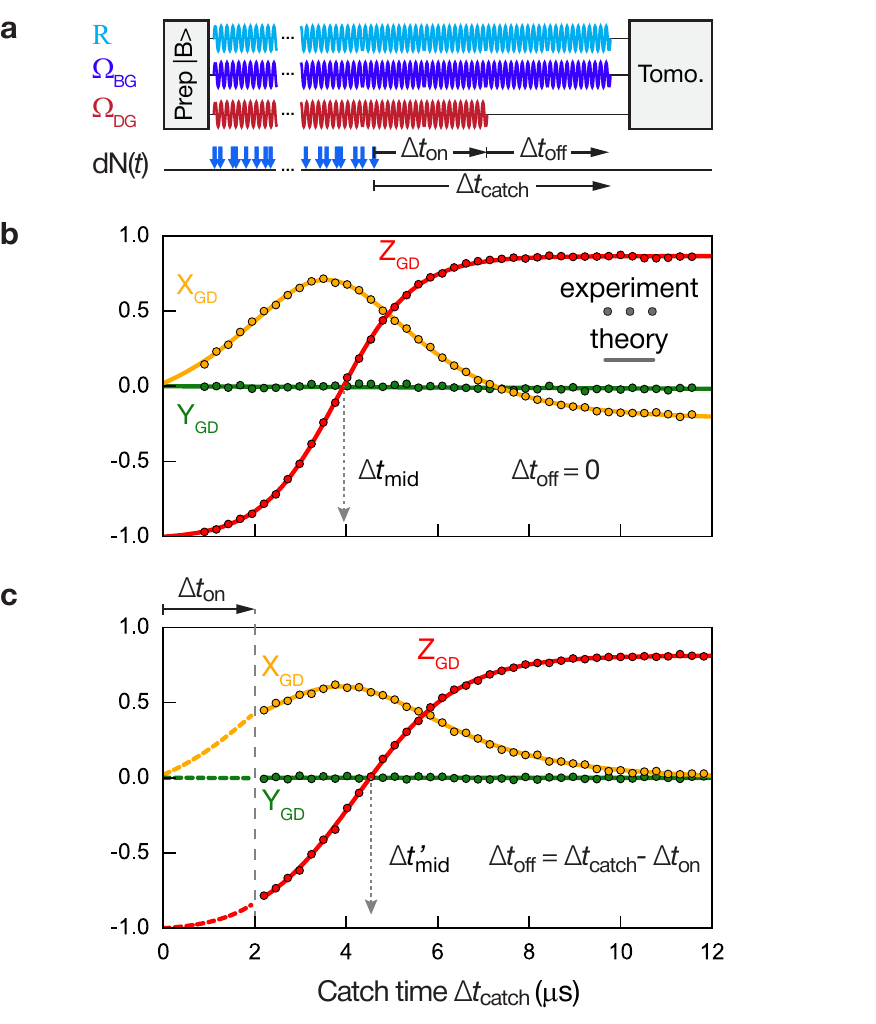}
\else
	\includegraphics[width=89mm]{3_catch.pdf}
\fi
\caption{
\label{fig:catch} 
\textbf{Catching the quantum jump mid-flight.} 
\textbf{a,} 
The atom is initially prepared in $\ket{\mathrm B}$.
The readout tone ($\mathrm R$) and atom Rabi drive $\Omega_\mathrm{BG}$ are turned on until the catch condition is fulfilled, consisting of the detection of a click followed by the  absence of click detections for a total time $\Delta t_\mathrm{catch}$.
The Rabi drive $\Omega_\mathrm{DG}$ starts with $\Omega_\mathrm{BG}$, but can be shut off prematurely, prior to the end of $\Delta  t_\mathrm{catch}$.
A tomography measurement is performed after $\Delta t_\mathrm{catch}$. 
\textbf{b \& c,}
Conditional tomography revealing the continuous, coherent, and, surprisingly, deterministic flight (when completed) of the quantum jump from $\ket{\mathrm{G}}$ to $\ket{\mathrm{D}}$.
The error bars are smaller than the size of the dots.
The mid-flight time $\Delta  t_{\mathrm{mid}}$ is defined by $Z_\mathrm{GD}=0$.
The jump proceeds even when $\Omega_\mathrm{DG}$ is turned off at the beginning of the flight (panel c), $\Delta  t_\mathrm{on}= 2\,\mathrm{\mu s}$. 
Data obtained from $6.8\times10^{6}$ experimental realizations.
Solid lines: theoretical prediction (Supplement Sec.~IIIA).
Dashed lines in panel c: theory curves for the $\Delta  t_\mathrm{on}$ interval, reproduced from panel b.
The data suggests that an advance-warning signal of the jump can be provided by a no-click period for catch time $\Delta t_\mathrm{catch} = \Delta t_\mathrm{mid}$, at which half of the jumps will complete. }
\end{centering}
\end{figure}

Having observed the quantum jumps in the free-running protocol, we  proceed to conditionally actuate the system control tones in order to tomographically reconstruct the time dynamics of the quantum jump from $\ket{\mathrm{G}}$ to $\ket{\mathrm{D}}$, see Fig.~3a.
Like previously, after initiating the atom in $\ket{\mathrm{B}}$, the FPGA controller continuously subjects the system to the atom drives ($\Omega_\mathrm{BG}$ and $\Omega_\mathrm{DG}$) and to the readout tone ($\mathrm R$).
However, in the event that the controller detects a single click followed by the complete absence of clicks for a total time $\Delta t_{\operatorname{catch}}$, the controller suspends all system drives, thus freezing the system evolution, and performs tomography, as explained in Methods. 
Note that in each realization, the tomography measurement yields a single +1 or -1 outcome, one bit of information for a single density matrix component. 
We also introduce a division of the duration $\Delta t_{\operatorname{catch}}$ into two phases, one lasting $\Delta t_\mathrm{on}$ during which $\Omega_\mathrm{DG}$ is left on and one lasting $\Delta t_\mathrm{off} = \Delta t_{\operatorname{catch}}- \Delta t_\mathrm{on}$ during which $\Omega_\mathrm{DG}$ is turned off. 
As we explain below, this has the purpose of demonstrating that the evolution of the jump is not simply due to the Rabi drive between $\ket{\mathrm{G}}$ and $\ket{\mathrm{D}}$.

In Fig.~3b, we show the dynamics of the jump mapped out in the full presence of the Rabi drive, $\Omega_\mathrm{GD}$, by setting $\Delta t_\mathrm{off}=0$. 
From $3.4\times10^{6}$ experimental realizations we reconstruct, as a function of $\Delta t_{\operatorname{catch}}$,  the quantum state, and present the evolution of the jump from $\ket{\mathrm{G}}$ to $\ket{\mathrm{D}}$ as the normalized, conditional GD tomogram (Methods).
For $\Delta t_{\operatorname{catch}} < 2\, \mathrm{\mu s}$, the atom is  predominantly detected in $\ket{\mathrm{G}}$ ($Z_\mathrm{GD} = -1$),  whereas for $\Delta t_{\operatorname{catch}}  > 10\, \mathrm{\mu s}$, it is predominantly detected in $\ket{\mathrm{D}}$  ($Z_\mathrm{GD} = +1$). 
Imperfections, mostly excitations to higher levels, reduce the maximum observed value to $Z_\mathrm{GD} = +0.9$ (Supplement Sec.~IIIB2). 
For intermediate no-click times, between $ \Delta  t_{\operatorname{catch}} =2 \, \mathrm{\mu s}$ and $\Delta t_{\operatorname{catch}} = 10\, \mathrm{\mu s}$, the state of the atom evolves  continuously and coherently from $\ket{\mathrm{G}}$ to $\ket{\mathrm{D}}$ ---  the  flight of the quantum jump. 
The time of mid flight,  $\Delta t_\mathrm{mid} \equiv 3.95\, \mathrm{\mu s}$, is markedly shorter than the Rabi period $2\pi/\Omega_\mathrm{DG} = 50\,  \mathrm{\mu s}$, and is given by the  function $ 
\Delta t_{\text{mid}}=\left(\frac{\Omega_\mathrm{BG}^{2}}{2\Gamma}\right)^{-1}\ln\left(\frac{\Omega_\mathrm{BG}^{2}}{\Omega_\mathrm{DG} \Gamma}+1\right)
$, in which $\Omega_\mathrm{DG}$ enters logarithmically (Supplement Sec.~IIA). 
The maximum coherence of the superposition, corresponding to $\sqrt{X_\mathrm{GD}^2 + Y_\mathrm{GD}^2}$, during the flight is $0.71 \pm 0.005$, see also Extended Data Fig.~\ref{fig:tomo_tmid}, quantitatively understood to be limited by several small imperfections (Supplement Sec.~IIIB2). 

Motivated by the quantum trajectory analysis, we fit the experimental data with
 $\mathrm{Z}_\text{GD}(\Delta t_{\operatorname{catch}} ) = a +b \tanh( \Delta t_{\operatorname{catch}}/\tau + c) $, 
$\mathrm{X}_\text{GD}(\Delta t_{\operatorname{catch}} ) = a' +b' \operatorname{sech}(\Delta t_{\operatorname{catch}}/\tau' + c') $, and $\mathrm{Y}_\text{GD}(\Delta t_{\operatorname{catch}} ) =  0$. 
We compare the fitted jump parameters ($a,a',b,b',c,c',\tau,\tau'$) to those 
calculated from the theory and numerical simulations using independently measured system characteristics, and find agreement at the percent level (Supplement Sec.~IIIA). 

By repeating the experiment with $\Delta t_\text{on} = 2\, \mathrm{\mu s}$, in Fig.~3c, we show that the jump proceeds even if the GD drive is shut off at the beginning of the no-click period.
The jump remains coherent and only differs from the previous case in a minor renormalization of the overall amplitude and timescale.
The mid-flight time of the jump, $\Delta t_\mathrm{mid}'$, is given by a modified formula (Supplement Sec.~IIA3). 
The results demonstrate that the role of  the Rabi drive $\Omega_\mathrm{DG}$ is to initiate the jump and provide a reference for the phase of its evolution\footnote{
A similar phase reference for a non-unitary, yet deterministic,  evolution induced by measurement was previously found in a different context in: N. Katz, M. Ansmann, R. C. Bialczak, E. Lucero, R. McDermott, M. Neeley, M. Steffen, E. M. Weig, A. N. Cleland, J. M. Martinis, and A. N. Korotkov, Science (New York, N.Y.) 312, 1498 (2006).
}. 
Note that the $\Delta t_\mathrm{catch} \gg \Delta t_\mathrm{mid}$ non-zero steady state value of $X_\mathrm{GD}$ in Fig.~3b is the result of the competition between the Rabi drive $\Omega_\mathrm{DG}$ and the effect of the measurement of $\ket{\mathrm{B}}$ (Supplement Sec.~IIA2).
This is confirmed in Fig.~3c, where $\Omega_\mathrm{DG} = 0$, and where there is no offset in the steady state value.  

The results of Fig.~\ref{fig:catch} demonstrate that despite the unpredictability  of the jumps from $\ket{\mathrm{G}}$ to $\ket{\mathrm{D}}$, they are preceded by an identical no-click record. While the jump starts at a random time and can be prematurely interrupted by a click, the deterministic nature of the uninterrupted flight comes as a surprise given the quantum fluctuations in the heterodyne record $I_\mathrm{rec}$ during the jump --- an island of predictability in a sea of uncertainty.

\begin{figure}
\begin{centering}
\ifjournal
	\includegraphics[width=89mm]{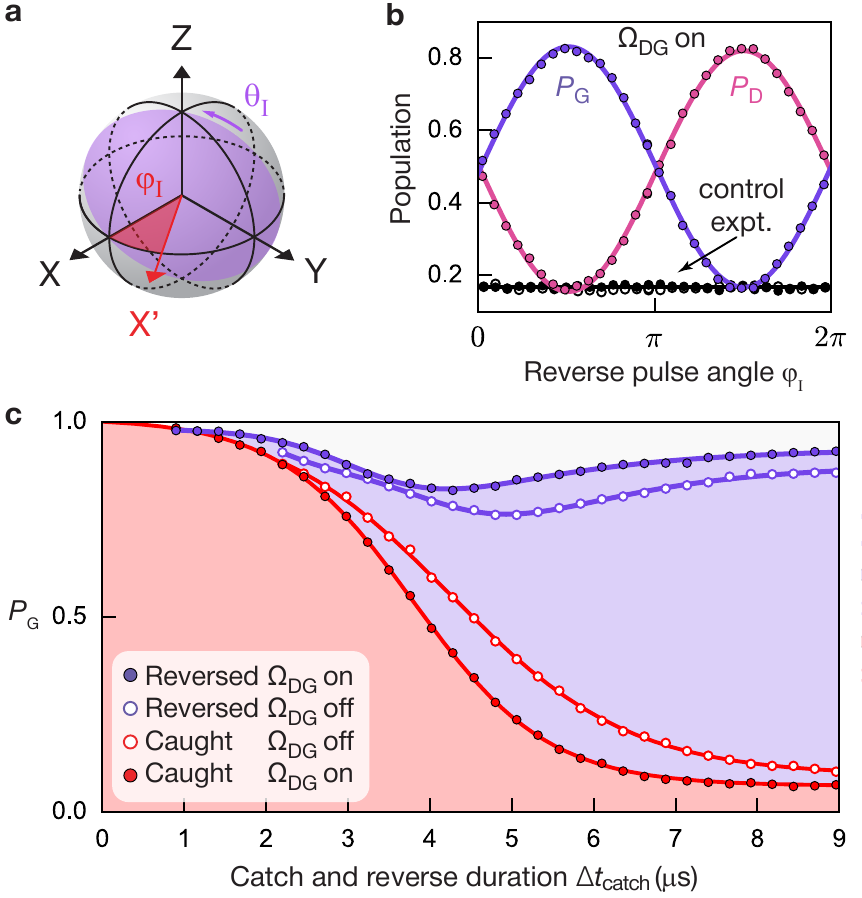}
\else
	\includegraphics[width=89mm]{4_reverse.pdf}
\fi
\caption{
\label{fig:reverse} 
\label{fig:Reversing-the-quantum}
\textbf{Reversing the quantum jump mid-flight.}
\textbf{a,} Bloch sphere of the GD manifold, showing the axis X' for the jump reversal, defined by the azimuthal angle $\varphi_\mathrm{I}$. The angle of the intervention pulse is $\theta_\mathrm{I}$.
\textbf{b,} 
Success probabilities $P_\mathrm G$ (purple) and $P_\mathrm D$ (orange) to reverse to $\ket{\mathrm G}$ and complete to $\ket{\mathrm D}$ the quantum jump mid-flight at $\Delta t_{\mathrm{catch}} = \Delta t_{\mathrm{mid}}$, with $\theta_\mathrm I = \pi/2$, in the presence of the Rabi drive $\Omega_\mathrm{DG}$.
The error bars are smaller than the size of the dots.
Black dots:  success probability for $\ket{\mathrm G}$ (closed dots) and $\ket{\mathrm D}$ (open dots) in a control experiment where intervention is applied at random times along the record, rather than at  $\Delta t_{\mathrm{catch}}$.
\textbf{c,}
Optimal success of reverse protocol (purple) as a function of $\Delta t_{\mathrm{catch}}$.  The FPGA controller is programmed with the optimal $\left\{\theta_{I}\left(\Delta t_{\mathrm{catch}}\right),\varphi_{I}\left(\Delta t_{\mathrm{catch}}\right)\right\}$.
Closed and open dots correspond to $\Delta t_\mathrm{on} = \Delta t_\mathrm{catch} $ and  $\Delta t_\mathrm{on} = 2\,\mathrm{\mu s} $, respectively. 
Red points show the corresponding open-loop (no intervention) results from Fig.~3b and~c. 
}
\end{centering}
\end{figure}

In Fig.~\ref{fig:reverse}b,  we show that by choosing $\Delta t_{\operatorname{catch}} = \Delta  t_\mathrm{mid}$ for the no-click period to serve as an advance warning signal, we reverse the quantum jump\footnote{
Reversal of quantum jumps have been theoretically considered in a different context, see H. Mabuchi and P. Zoller, Phys. Rev. Lett. 76, 3108 (1996).}
in the presence of  $\Omega_\mathrm{DG}$, confirming its coherence; the same result is found when $\Omega_\mathrm{DG}$ is off,  see Extended Data Fig.~4.
The reverse pulse characteristics are defined in Fig.~\ref{fig:reverse}a. 
For $\varphi_\mathrm{I} = \pi/2$, our feedback protocol succeeds in reversing the jump to $\ket{\mathrm{G}}$ with  $83.1\%\pm0.3\%$ fidelity, while for $\varphi_\mathrm{I} = 3\pi/2$, the protocol completes the jump to $\ket{\mathrm{D}}$, with  $82.0\%\pm0.3\%$ fidelity. 
In a control experiment, we repeat the protocol by applying the reverse pulse at random times, rather than those determined by the advance warning signal. 
Without the advance warning signal, the measured populations only reflect those of the ensemble average.

In a final experiment, we programmed the controller with the optimal reverse pulse parameters $\left\{\theta_{I}\left(\Delta  t_{\mathrm{catch}}\right),\varphi_{I}\left(\Delta  t_{\mathrm{catch}}\right)\right\}$, and as shown in Fig.~\ref{fig:reverse}c, we measured the success of the reverse protocol as a function of the catch time, $\Delta t_\mathrm{catch}$.
The closed/open dots indicate the results for $\Omega_\mathrm{DG}$ on/off, while the  solid curves are theory fits motivated by the exact analytic expressions (Supplement Sec.~IIIA).
The complementary red dots and curves reproduce the open-loop results of Fig.~\ref{fig:catch} for comparison.

From the experimental results of Fig.~2a one can infer, consistent with Bohr's initial intuition and the original ion experiments, that quantum jumps are random and discrete. 
Yet, the results of Fig.~3 support a contrary view, consistent with that of Schrödinger: the evolution of the jump is coherent and continuous. 
Noting the difference in time scales in the two figures, we interpret the coexistence of these seemingly opposed point of views as a unification of the discreteness of countable events like jumps  with the continuity of the deterministic Schrödinger’s equation.
Furthermore, although all $6.8\times10^6$ recorded jumps (Fig.~3) are entirely independent of one another and stochastic in their initiation and termination, the tomographic measurements as a function of $\Delta t_\mathrm{catch}$ explicitly show that all jump evolutions follow an essentially identical, predetermined path in Hilbert space — not a randomly chosen one — and, in this sense, they are deterministic.  These results are further corroborated by  the reversal experiments shown in Fig.~4,  which exploit the continuous, coherent, and deterministic nature of the jump evolution and critically hinge on priori knowledge of the Hilbert space path. 
With this knowledge ignored in the control experiment of Fig. 4b,  failure of the reversal is observed. 

In conclusion, these experiments revealing the coherence of the jump, promote the view that a single quantum system under efficient, continuous observation is characterized by a time-dependent state vector inferred from the record of previous measurement outcomes, and whose meaning is that of an objective, generalized degree of freedom.
The knowledge of the system on short timescales is not incompatible with an unpredictable switching behavior on long time scales. 
The excellent agreement between experiment and theory including known experimental imperfections  (Supplement Sec.~IIIA) thus provides support to the modern quantum trajectory theory and its reliability for predicting the performance of real-time intervention techniques in the control of single quantum systems.

\footnotesize
\textbf{Acknowledgments}\enspace
Z.K.M. acknowledges fruitful discussion with S.M. Girvin, H.M. Wiseman, K. M{\o}lmer, N. Ofek,  V.V. Albert, and M.P. Silveri.  V.V. Albert addressed one aspect of the Lindblad theoretical modeling regarding the waiting-time distribution. 
Facilities use was supported by the  Yale Institute for Nanoscience and Quantum Engineering (YINQE), the Yale SEAS cleanroom, and NSF MRSEC DMR 1119826. This research
was supported by ARO under Grant No. W911NF-14-1-0011.
R.G.J. and H.J.C. acknowledge the support of the Marsden Fund Council from Government funding, administered by the Royal Society of New Zealand under Contract No UOA1328.

\textbf{Author Contributions}\enspace
Z.K.M. initiated and performed the experiment, designed the sample, analyzed the data, and carried out the initial theoretical and numerical modeling of the experiment. 
Z.K.M. conceived the experiment based on theoretical predictions by H.J.C.
H.J.C. and R.G.J. performed the presented theoretical modeling and numerical simulations.
S.O.M. contributed to the experimental setup and design of the device, and with S.S. to its fabrication.
P.R. and R.J.S. assisted with the FPGA.  
M.M. contributed theoretical support, and M.H.D. supervised the project.
Z.K.M. and M.H.D. wrote the manuscript, and H.J.C. contributed the theoretical supplement. 
All authors provided suggestions for the experiment, discussed the results and contributed to the manuscript.

\textbf{Correspondence}\enspace 
Correspondence and requests for materials should be addressed to Z.K. Minev~(email: zlatko.minev@aya.yale.edu) and M.H. Devoret~(email: michel.devoret@yale.edu)

\normalsize

\ifUSEMULTIBIB
	\nocite{apsrev41Control}
	\bibliographystyle{apsrev4-1}
	\bibliography{mycontrol,biblio}
\fi

\clearpage
\normalsize

\onecolumngrid
\section*{Methods}
\twocolumngrid

\makeatletter
\renewcommand{\fnum@figure}{\textbf{Extended Data Figure~\thefigure}}
\renewcommand{\fnum@table}{\textbf{Extended Data Table~\thetable}}
\makeatother

\setcounter{figure}{0}
\setcounter{table}{0}

\textbf{Monitoring quantum jumps}
\label{sec:What_Are_Quantum_Jumps}

Here, we aim to briefly explain  how  the GD dynamics is monitored and when do we conclude that a quantum jump has occurred.

\textit{Monitoring the GD manifold through B de-excitations.}
The state of the atom within the GD manifold is monitored indirectly, by measuring the rate of de-excitations from the ancillary state $\ket{\mathrm{B}}$, while the G to B excitation tone $\Omega_\mathrm{BG}$ is applied.
As explained below, the monitoring scheme is such that when the atom is in the dark state, $\ket{\mathrm{D}}$, the rate of de-excitations from $\ket{\mathrm{B}}$ to $\ket{\mathrm{G}} $ is zero.
Conversely, when the atom is in  $\ket{\mathrm{G}}$, the rate  is non-zero.
Henceforth, we will  refer to a de-excitation from $\ket{\mathrm{B}}$ to $\ket{\mathrm{G}}$  simply as a de-excitation.
In summary,  when the rate of de-excitations for a measurement segment is zero, $\ket{\mathrm{D}}$ is assigned to it; otherwise, $\ket{\mathrm{G}}$ or $\ket{\mathrm{B}}$ is assigned (see IQ filter section of Methods).
The rate can be monitored by either a direct or indirect method, as explained further below.  

\textit{Quantum jumps.} 
Sections of the (continuous) measurement record are converted into state assignments, as discussed above, such as B, G, or D.
In the experiment, long sequences of such measurements yield the same result, i.e., GGG\ldots or DDD\ldots 
When the string of results suddenly switches its value, we say that a quantum jump has occurred.\citeMethods{Cook1988}

\textit{Source of the difference for the de-excitation rates.}
The rate of de-excitations is zero when the atom is in $\ket{\mathrm{D}}$ because the V-shape level structure forbids any direct DB transitions; hence,  $\ket{\mathrm{B}}$ cannot be excited from $\ket{\mathrm{D}}$. 
Conversely, when the atom is in $\ket{\mathrm{G}}$, the Rabi drive $\Omega_\mathrm{BG}$ can excite the atom to $\ket{\mathrm{B}}$. Since this ancillary state is effectively short-lived, it almost immediately de-excites back to $\ket{\mathrm{G}}$. 
Note that in this explanation we neglect parasitic transitions to higher excited states, which are considered in the Supplement.

\textit{De-excitation detection: direct or indirect.}
A de-excitation can be detected by a direct or, alternatively, indirect method. 
For atomic experiments, direct detection is a natural choice. 
The photon emitted by the atom during the de-excitation, carrying away the energy once stored in $\ket{\mathrm{B}}$,  is collected and destructively absorbed in the sensor of a photodetecting measurement apparatus, which produces a ``click'' signal (in practice, a current or voltage pulse).
Unfortunately, unavoidable imperfections and detector non-idealities  prohibit this method for the continuous detection of nearly every single de-excitation (Supplement Sec.~III).
Alternatively, one can use an indirect monitoring method.
In our experiment, instead of detecting the emitted photon, we  detect the  de-excitation by monitoring the $\ket{\mathrm{B}}$ population through an ancillary degree of freedom, the readout cavity, coupled to the atom. 

\textit{Indirect (dispersive) detection.}
The readout cavity frequency depends on the state of the atom. 
When the atom is in $\ket{\mathrm{B}}$, the readout cavity frequency shifts down by more than a cavity linewidth. 
The cavity frequency, and hence the $\ket{\mathrm{B}}$ population of the atom, is probed by a continuous readout tone applied at the $\ket{\mathrm{B}}$-cavity frequency. 
When the atom is in $\ket{\mathrm{B}}$, the probe tone is resonant and fills the cavity with a large number of photons, $\bar{n}$. Otherwise, when the atom is \textit{not}-in-$\ket{\mathrm{B}}$, the probe tone is far off resonant and the cavity is empty of photons. 
Choosing $\bar{n}\gg1$ makes a change in the $\ket{\mathrm{B}}$ occupancy conspicuous, and hence a de-excitation, $\ket{\mathrm{B}}$ to not-$\ket{\mathrm{B}}$, is readily observed, even in the presence of measurement inefficiencies and imperfections.
As explained in Sec.~IIIC of the Supplement, this indirect dispersive method in effect increases the signal-to-noise ratio (SNR) and de-excitation detection efficiency. 
Another notable difference between the direct and indirect method is that in the indirect method the atom fully but shortly occupies  $\ket{\mathrm{B}}$  before de-exciting to  $\ket{\mathrm{G}}$, while in the direct scheme the  probability amplitude to be in $\ket{\mathrm{B}}$  is never appreciable before a de-excitation, see Sec.~II of the Supplementary Information.
In other words, in the direct monitoring scheme, there are explicitly two sets of quantum jumps: the BG and DG ones. The BG ones occurs much faster and are nested within the DG jumps. The fast dynamics of these ``inner'' jumps is used to interrogate the dynamics of ``outer,'' DG jumps.

\textbf{Setup of the Experiment}

\textit{Setup and signals.} Our experiments were carried out in a cryogen-free dilution refrigerator (\textit{Oxford Triton 200}). The cavity and JPC\citeMethods{Bergeal2010} were shielded from stray magnetic fields by a cryogenic $\mu$-metal (\textit{Amumetal A4K}) shield. Our input-output cryogenic setup is nearly identical to that described in Ref.~\onlinecite{Ofek2016}, aside from the differences evident in the schematic of our setup (see Fig.~1b and Methods) or described in the following.

The control tones depicted in Fig.~1 were each generated from  individual microwave generators ($\Omega_\mathrm{DG}$ and $\Omega_\mathrm{B0}$: \textit{Agilent N5183A}; readout cavity tone R and $\Omega_\mathrm{B1}$: \textit{Vaunix LabBrick LMS-103-13} and LMS-802-13, respectively). To achieve IQ control, the generated tones were mixed (\textit{Marki Microwave Mixers IQ-0618LXP} for the cavity and \textit{IQ-0307LXP} for $\Omega_\mathrm{B0}, \Omega_\mathrm{B1}$, and $\Omega_\mathrm{DG}$) with intermediate-frequency (IF) signals synthesized by the 16 bit digital-to-analog converters (DACs) of the integrated FPGA controller system (\textit{Innovative Integration VPXI-ePC}). Prior to mixing, each analog output was filtered by a $50\,\Omega$ low pass filter (\textit{Mini-Circuits BLP-300+}) and attenuated by a minimum of 10\,dB.
The radio-frequency (RF) output was amplified at room temperature (\textit{MiniCircuits ZVA-183-S+}) and filtered by \textit{Mini-Circuits} coaxial bandpass filters.
The output signal was further pulse modulated by the FPGA with high isolation SPST switches (\textit{Analog Device HMC-C019}), which provided additional 80\,dB isolation when the control drives were turned off.
The signals were subsequently routed to the input lines of the refrigerator, whose details were described in Refs.~\onlinecite{Ofek2016} and~\onlineciteMethods{Minev2016}.

At room temperature, following the cryogenic high-electron mobility amplifier (HEMT;  \textit{Low Noise Factory LNF-LNC7\_10A}), the signal were amplified by 28\,dB (\textit{Miteq AFS3-00101200-35-ULN}) before being mixed down (\textit{Marki image reject double-balanced mixer IRW-0618}) to an intermediate frequency (IF) of 50\,MHz, where they were band-pass filtered (\textit{Mini-Circuits SIF-50+}) and further amplified by a cascaded preamplifier (\textit{Stanford Research Systems SR445A}), before finally digitization by the FPGA analog-to-digital converters (ADC).

\textbf{Atom-cavity implementation}

The superconducting artificial atom consisted of two coupled transmon qubits fabricated on a 2.9~mm-by-7~mm double-side-polished  c-plane sapphire wafer with the Al/Al$\text{O}_\text{x}$/Al bridge-free electron-beam lithography technique\citeMethods{Lecocq2011-bridge-free, Rigetti2009}. 
The first transmon (B) was aligned with the electric field of the fundamental $\mathrm{TE}_\mathrm{101}$ mode of an aluminum rectangular cavity (alloy 6061; dimensions: 5.08~mm by 35.5~mm by 17.8~mm), while the second transmon (D) was oriented perpendicular to the first and  positioned $170\,\mathrm{\mu m}$ adjacent to it.
The inductance of the Josephson junction of each transmon (9~nH for both B and D), the placement and dimensions of each transmon, and the geometry of the cavity were designed and optimized using finite-element electromagnetic analysis and the energy-participation-ratio (EPR) method\citeMethods{Minev2018-EPR}. The analysis also verified that the coupling  between the two qubits is described by the Hamiltonian $\hat{H}_\mathrm{int} = -\chi_\mathrm{DB} \hat{n}_\mathrm{B} \otimes  \hat{n}_\mathrm{D}$, where $\hat{n}_\mathrm{B/D}$ is the photon number operator of the B/D qubit, and  $\chi_\mathrm{DB}$ is the cross-Kerr frequency. 

The measured frequency and anharmonicity of the D qubit were $\omega_\mathrm{D}/2\pi = 4845.255\,\mathrm{MHz}$ and  $\alpha_\mathrm{DG}/2\pi = 152\,\mathrm{MHz}  $, respectively, while those of the B qubit were $\omega_\mathrm{B}/2\pi = 5570.349 \,\mathrm{MHz}$ and $\alpha_\mathrm{BG}/2\pi = 195 \,\mathrm{MHz}$, respectively. 
The cross-Kerr coupling was $\chi_\mathrm{DB} /2\pi = 61\,\mathrm{MHz}$.
The relaxation time of  $\ket{\mathrm{B}}$  was $T_\mathrm{1}^{\mathrm{B}} =28 \pm 2\, \mathrm{\mu s}$, limited by the Purcell effect by design, while its Ramsey coherence time was  $T_\mathrm{2R}^\mathrm{B} =  18\pm1\, \mathrm{\mu s}$.
The remaining parameters of the system are provided in the main text.

\textbf{Atom and cavity drives} 

In all experiments, the following drive parameters were used: 
The DG Rabi drive, $\Omega_\mathrm{DG}$, was applied 275\,kHz below $\omega_\mathrm{D}$, to account for the Stark shift of the cavity. 
The BG drive, $\Omega_\mathrm{BG}$, was realized as a bi-chromatic tone in order to unselectively address the BG transition, which was broadened and Stark shifted due to the coupling between $\ket{\mathrm{B}}$ and the readout cavity. 
Specifically, we address transitions from $\ket{\mathrm{G}}$ to $\ket{\mathrm{B}}$ with a Rabi drive $\Omega_\mathrm{B0}/2\pi = 1.20\pm0.01\,$MHz at frequency $\omega_\mathrm{BG}$, whereas transitions from $\ket{\mathrm{B}}$ to $\ket{\mathrm{G}}$ are addressed with a Rabi drive $\Omega_\mathrm{B1}/2\pi = 0.60\pm0.01\,$MHz tuned 30\,MHz below $\omega_\mathrm{BG}$. This bi-chromatic scheme provided the ability to tune the up-click and down-click rates independently, but otherwise essentially functioned as an incoherent broad-band source. 

\textbf{IQ filter} 

To mitigate the effects of imperfections in the atom readout scheme in extracting a $\ket{\mathrm{B}}$/not-$\ket{\mathrm{B}}$ result, we applied a two-point, hysteretic IQ filter, implemented on the FPGA controller in real time. 
The filter is realized by comparing the present quadrature record values $ \left\{ I_\mathrm{rec}, Q_\mathrm{rec}\right\} $, with three thresholds ($I_\mathrm{B}, I_{\bar{\mathrm{B}}},$ and $Q_\mathrm{B}$) as summarized in Extended Data Table~\ref{table:iq-filter-logic}.
	
The filter and thresholds were selected to provide a best estimate of the time of a click, operationally understood as a change in the filter output from $\ket{\mathrm{B}}$ to not-$\ket{\mathrm{B}}$.
The $I_{{\mathrm{B}}}$ and  $I_{\bar{\mathrm{B}}}$ thresholds were chosen 1.5 standard deviations away from the I-quadrature mean of the $\ket{\mathrm{B}}$ and not-$\ket{\mathrm{B}}$  distributions, respectively.
The  $Q_{{\mathrm{B}}}$ threshold was chosen 3 standard deviations away from the Q-quadrature mean. Higher excited states of the atom were selected out by $Q_\mathrm{rec} $ values exceeding the $Q_\mathrm{B}$ threshold.

\textbf{Tomography} 

At the end of each experimental realization, we performed one of 15 rotation sequences on the atom that transferred information about one component of the density matrix, $\hat{\rho}_a$, to the population of $\ket{\mathrm{B}}$, which was measured with a 600\,ns square pulse on the readout cavity.
Pulses were calibrated with a combination of Rabi, derivative removal via adiabatic gate (DRAG)\citeMethods{JChow2010-DRAG}, All-XY\citeMethods{Reed2013}, and amplitude pulse train sequences\citeMethods{Bylander2011}. 
The readout signal was demodulated with the appropriate digital filter function required to realize temporal mode matching\citeMethods{Eichler2012-itinerant-entanglement}.
To remove the effect of potential systematic offset errors in the readout signal, we subtracted the measurement results of operator components of  $\hat{\rho}_a$ and their opposites.
From the measurement results of this protocol, we reconstructed the density matrix $\hat{\rho}_a$, and subsequently parametrized it in the useful form
$$\hat{\rho}_a = \begin{pmatrix}\frac{N}{2}\left(1-Z_{\mathrm{GD}}\right) & \frac{N}{2}\left(X_{\mathrm{GD}}+i Y_{\mathrm{GD}}\right) & R_{\mathrm{BG}}+i I_{\mathrm{BG}}\\
\frac{N}{2}\left(X_{\mathrm{ GD}}-i Y_{\mathrm{GD}}\right) & \frac{N}{2}\left(1+Z_{\mathrm{GD}}\right) & R_{\mathrm{BD}}+i I_{\mathrm{BD}}\\
R_{\mathrm{BG}}-i I_{\mathrm{BG}} & R_{\mathrm{BD}}-i I_{\mathrm{BD}} & 1-N
\end{pmatrix}, 
$$ where $X_\mathrm{GD}, Y_\mathrm{GD},$ and $Z_\mathrm{GD}$ are the Bloch vector components of the GD manifold, $N$ is the total population of  the $\ket{\mathrm{G}}$ and $\ket{\mathrm{D}}$ states, while $R_{\mathrm{BG}},  R_{\mathrm{BD}}, I_{\mathrm{BG}}$ and $I_{\mathrm{BD}}$ are the coherences associated with $\ket{\mathrm{B}}$, relative to the GD manifold. 
The measured population in $\ket{\mathrm{B}}$, $1-N$, remains below 0.03 during the quantum jump, see Extended Data Fig.~\ref{fig:reverseDGoff}.
Tomographic reconstruction was calibrated and verified by preparing Clifford states, accounting for the readout fidelity of 97\%.

\textbf{Control flow of the experiment}
 
\begin{figure}[!ht]
\begin{centering}
\hspace*{-1cm} \ifjournal
	\includegraphics[width=189mm]{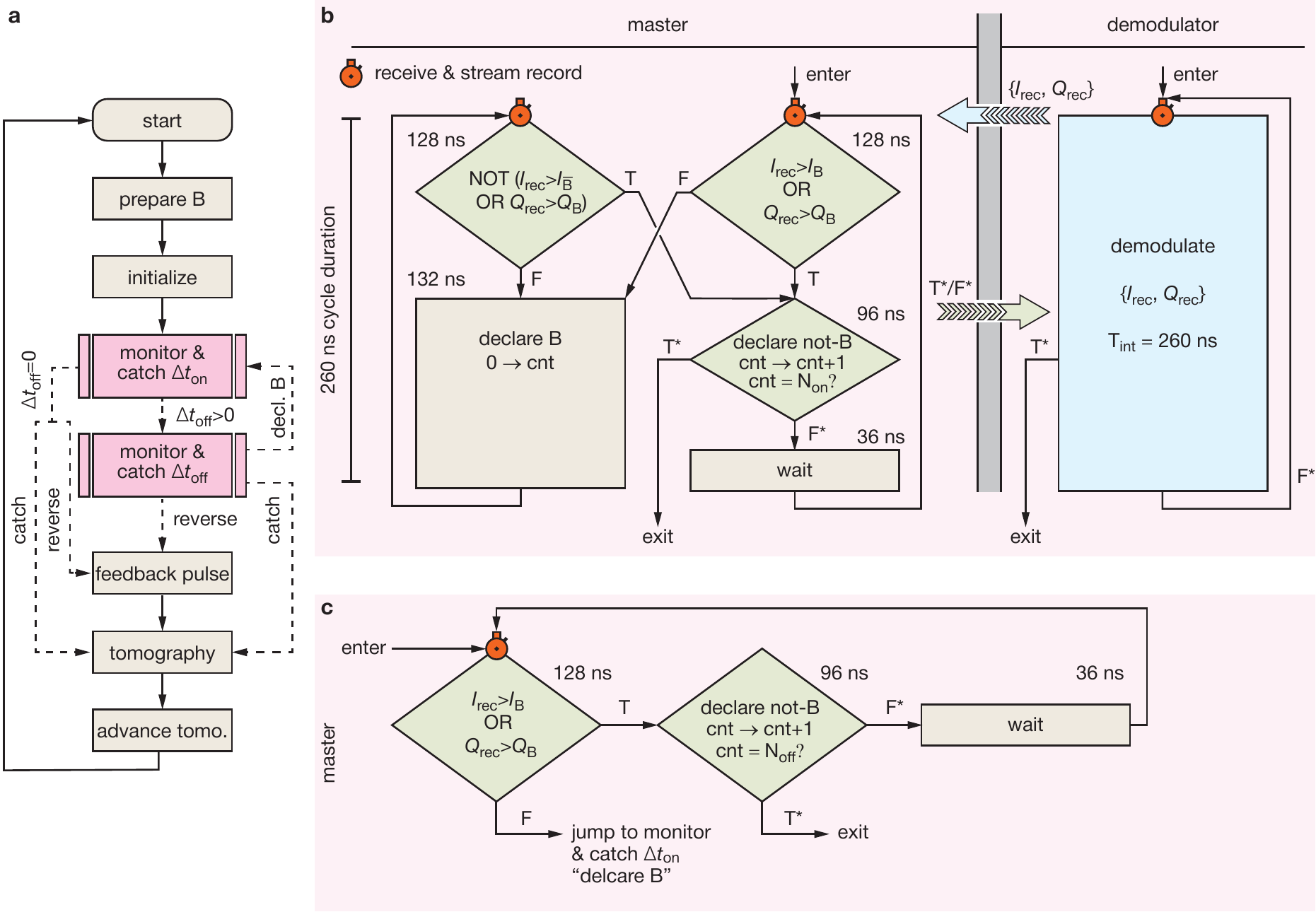} 
\else
	\includegraphics[width=189mm]{experiment_flow.pdf} 
\fi
\par
\caption{ \label{fig:experiment-flow}
\textbf{Control flow of the experiment.} 
\textbf{a,} 
Flowchart illustrating the control flow of the catch and reverse experiments, whose results are shown in Figs.~3 and~4.  See Methods for the description of each block.
\textbf{b,} 
Flowchart  of the master and demodulator modules chiefly involved in the ``monitor and catch $\Delta t _\mathrm{on}$'' routine. The modules execute concurrently and share data synchronously, as discussed in Methods. 
\textbf{c,}
Flowchart of the processing involved in the master module  of the `monitor and catch~$\Delta t _\mathrm{off}$'' routine; see Methods. 
}
\end{centering}
\end{figure}

\begin{figure}[!ht]
\begin{centering}
\ifjournal
	\includegraphics[width=105mm]{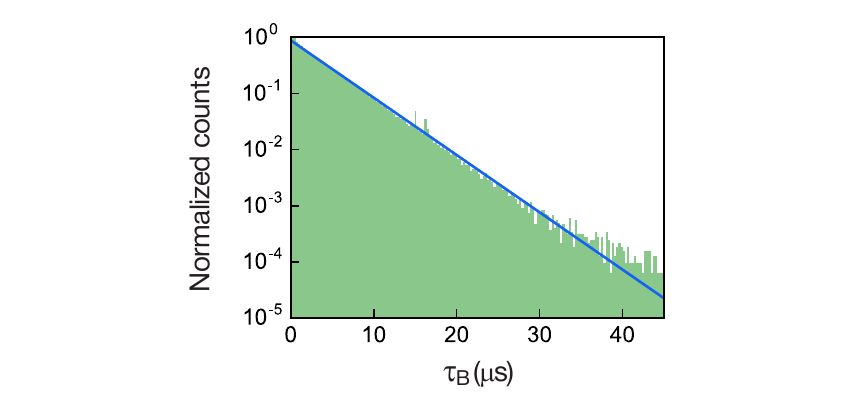} 
\else
	\includegraphics[width=105mm]{B_waiting_time.pdf} 
\fi
\caption{ \label{fig:B-wait-time}
\textbf{Waiting time to switch from a $\B$ to not-$\B$ state assignment result.}  
Semi-log plot of the histogram (shaded green) of the duration of times corresponding to $\B$-measurement results, $\tau_{\operatorname{B}}$, for 3.2 s of continuous data of the type shown in Fig.~2a.
Solid line is an exponential fit, which yields a $4.2\pm0.03\,\mathrm{\mu s}$ time constant.
}
\end{centering}
\end{figure}

\begin{figure}[!ht]
\begin{centering}
\ifjournal
	\includegraphics[width=105mm]{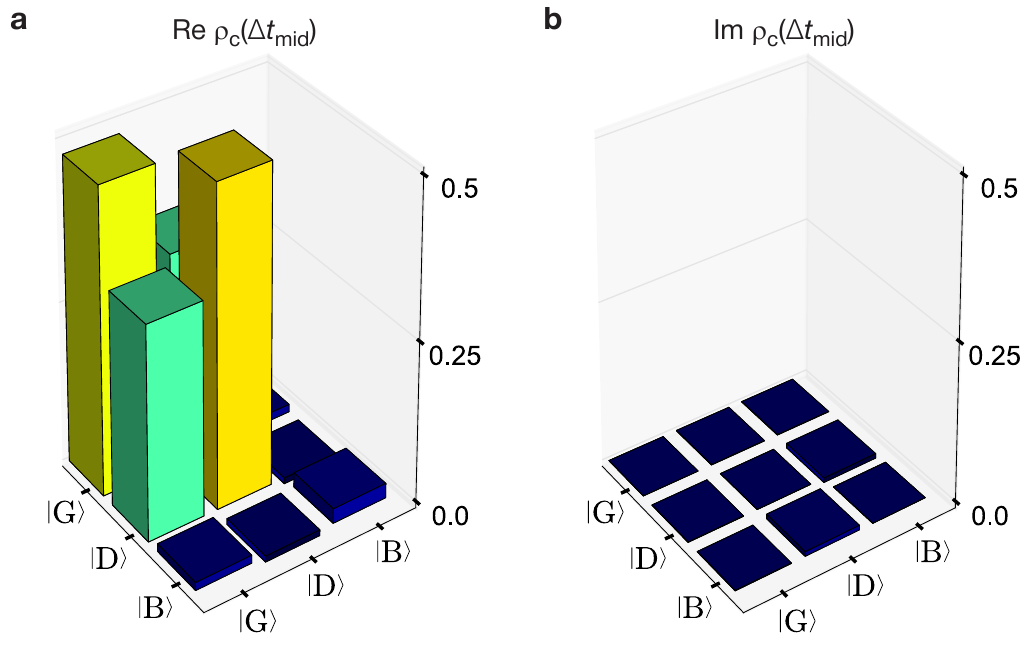} 
\else
	\includegraphics[width=105mm]{tomogram_tmid.pdf} 
\fi
\caption{ \label{fig:tomo_tmid}
\textbf{Mid-flight tomogram.}  
The plots show the real (a) and imaginary (b) parts of the conditional density matrix, $\rho_\mathrm{c}$, at the mid flight of the quantum jump ($\Delta t_\mathrm{catch} = \Delta t_\mathrm{mid}$), in the presence of the Rabi drive from $\ket{\mathrm{G}}$ to $\ket{\mathrm{D}}$ ($\Delta t_\mathrm{off} = 0$).
The population of the $\ket{\mathrm{B}}$ state is 0.023, and the magnitude of all imaginary components is less than 0.007. 
}
\end{centering}
\end{figure}

\begin{figure}[!ht]
\begin{centering}
\ifjournal
	\includegraphics[width=89mm]{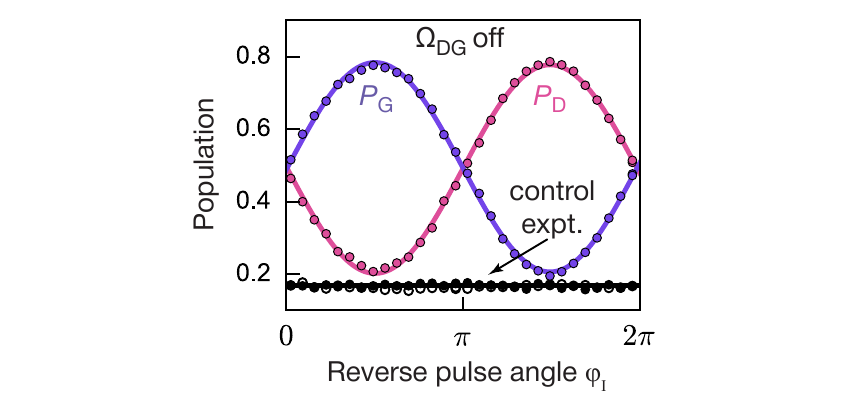} 
\else
	\includegraphics[width=89mm]{reverse_DG_off_tmid.pdf} 
\fi
\caption{ \label{fig:reverseDGoff}
\textbf{Reversing the quantum jump mid-flight in the absence of $\mathbf{\Omega_\mathrm{\textbf{DG}}}$.}  
Success probabilities $P_\mathrm G$ (purple) and $P_\mathrm D$ (orange) to reverse to $\ket{\mathrm G}$ and complete to $\ket{\mathrm D}$ the quantum jump mid-flight at $\Delta t_{\mathrm{catch}} = \Delta t'_{\mathrm{mid}}$, defined in Fig.~3b, in the absence of the Rabi drive $\Omega_\mathrm{DG}$, where $\Delta t_\mathrm{on} = 2\, \mathrm{\mu s}$ and $\theta_\mathrm I = \pi/2$.
Black dots:  success probability for $\ket{\mathrm G}$ (closed dots) and $\ket{\mathrm D}$ (open dots) for the control experiment where the intervention is applied at random times, see Fig.~4b.
}
\end{centering}
\end{figure}

\label{sec:experiment-flow}
A diagrammatic representation of the control flow of the experiment is illustrated in Extended Data Figure~\ref{fig:experiment-flow}a, whose elements are briefly described in the following.
``Start'': FPGA controller resets its internal memory registers to zero\cite{Ofek2016}$^,$ \citeMethods{Liu2016Thesis}, including the no-click counter ``cnt,'' defined below. 
``Prepare~B'': controller deterministically prepares the atom in $\ket{\mathrm{B}}$, a maximally conservative initial state, with measurement-based feedback\citeMethods{Riste2012-qubit-measure-reset}.
``Initialize'':  controller turns on the atom ($\Omega_\mathrm{BG}$ and~$\Omega_\mathrm{DG}$) and cavity ($\mathrm{R}$) drives and begins demodulation.
``Monitor and catch~$\Delta t_\mathrm{on}$'': with all drives on ($\Omega_\mathrm{BG}, \Omega_\mathrm{DG}$, and $\mathrm{R}$), the controller actively monitors the cavity  output signal until it detects no-clicks for duration $\Delta t_\mathrm{on}$, as described in panel (b), whereafter the controller proceeds to ``monitor and catch $\Delta t_\mathrm{off}$''  in the case that $\Delta t_\mathrm{off}>0$; otherwise, for $\Delta t_\mathrm{off}=0$, the controller proceeds to ``tomography''  (``feedback pulse'') for the catch (reverse) protocol.
``Monitor and catch~$\Delta t_\mathrm{off}$'': with the Rabi drive $\Omega_\mathrm{DG}$ off, while keeping the drives~$\Omega_\mathrm{BG}$ and~R on, the controller continues to monitor the output signal. The controller exits the routine only in one of two events: i) if it detects a click, in which case it proceeds to the ``declare B'' step of the ``monitor and catch~$\Delta t _\mathrm{on}$'' routine, or ii) if no further clicks are detected for the entirety of the pre-defined duration $\Delta t_\mathrm{off}$, in which case the controller advances to  the ``tomography'' (``feedback pulse'')  routine, when programmed for the catch (reverse) protocol.
``Feedback~pulse'': with all continuous-wave drives turned off, the controller performs a pulse on the DG transition of the atom, defined by the two angles $\left\{\theta_{I}\left(\Delta t_{\mathrm{catch}}\right),\varphi_{I}\left(\Delta t_{\mathrm{catch}}\right)\right\}$.
``Tomography'':  controller performs  next-in-order tomography sequence (see Tomography section above) while the demodulator finishes processing the final data in its pipeline. 
``Advance~tomo.'': tomography sequence counter is incremented; after a $50~\mathrm{\mu s}$ delay, the next realization of the experiment is started.

The concurrent-programming control flow of the``monitor and catch $\Delta t _\mathrm{on}$'' block is illustrated in Extended Data Fig.~\ref{fig:experiment-flow}b; specifically, the master and demodulator modules of the controller and synchronous sharing of data between them is depicted.
The FPGA demodulator outputs a pair of 16 bit signed integers, $\left\{ I_\mathrm{rec}, Q_\mathrm{rec}  \right\}$, every $T_\mathrm{int} = 260$~ns, which is routed to the master module, as depicted by the large left-pointing arrow (top).
The master module implements the IQ filter (see IQ filter section above) and tracks the number of consecutive not-$\ket{\mathrm{B}}$ measurement results with the counter ``cnt.'' The counter thus keeps track of the no-click time elapsed since the last click, which is understood as a change in the measurement result from $\ket{\mathrm{B}}$ to not-$\ket{\mathrm{B}}$.
When the counter reaches the critical value $N_\mathrm{on}$, corresponding to~$\Delta t_\mathrm{on}$, the master and demodulator modules synchronously exit the current routine, see the~T* branch of the ``declare not-B'' decision block.
Until this condition is fulfilled (F*), the two modules proceed within the current routine as depicted by the black flowlines.
To minimize latency and maximize computation throughput, the master and demodulator were designed to be independent sequential processes running concurrently on the FPGA controller,  communicating strictly through synchronous message passing, which imposed stringent synchronization and execution time constraints.
All master inter-module logic was constrained to run at a 260~ns cycle, the start of which necessarily was imposed to coincide with a ``receive \& stream record'' operation, here, denoted by the stopwatch. In other words, this imposed the algorithmic constraint that all flowchart paths staring at a stopwatch and ending in a stopwatch, itself or other, were constrained to a 260\,ns execution timing. A second key timing constraint was imposed by the time required to propagate signals between the different FPGA cards, which corresponded to a minimum branching-instruction duration of 76~ns.

The corresponding demodulation-module flowchart is identical to that shown of panel (b); hence, it is not shown.
This routine functions in following manner:
If a~$\ket{\mathrm{B}}$  outcome is detected, the controller jumps to the ``declare B'' block of the monitor \& catch $\Delta t_\mathrm{on}$ routine;
otherwise, when only not-$\ket{\mathrm{B}}$ outcomes are observed, and the counter reaches the critical value $N_\mathrm{off}$, corresponding to $\Delta t_\mathrm{catch} = \Delta t_\mathrm{on} + \Delta t_\mathrm{off}$, the controller exits the routine.

\footnotesize
\textbf{Data availability}\enspace
The data that support the findings of this study are available from the corresponding authors on reasonable request.
\normalsize

\clearpage
\onecolumngrid
\section*{Extended Data Figures} 

\begin{table*}[!h]
\begin{center}
	\begin{tabular}{r|c|c|c}
	Input: & $Q_{\mathrm{rec}}\geq Q_{\mathrm{B}}$ or $I_{\mathrm{rec}}> I_{\mathrm{B}}$ & $Q_{\mathrm{rec}}<Q_{\mathrm{B}}$ and $I_{\mathrm{rec}}<I_{\bar{\mathrm{B}}}$ & $Q_{\mathrm{rec}}<Q_{\mathrm{B}}$ and $I_{\bar{\mathrm{B}}}\leq I_{\mathrm{rec}} \leq I_{\mathrm{B}}$\tabularnewline
	\hline 
	Output: & $\ket{\mathrm{B}}$ & not-$\ket{\mathrm{B}}$ & previous\tabularnewline
	\end{tabular}
	\caption{\label{table:iq-filter-logic} Input-output table summarizing the behavior  of the IQ filter implemented on the FPGA controller.	}
\end{center}
\end{table*}

\begin{table}[!ht]
\begin{centering}
\addtolength{\tabcolsep}{2pt} \begin{tabular}{cc>{\raggedright}p{0.75\columnwidth}}
\hline 
\textbf{Symbol} & \textbf{Value} & \textbf{Description}\tabularnewline
\hline 
$\Gamma^{-1}$ & $\approx8.8$\,ns & Effective measurement time of $\ket{\mathrm{B}}$, approximately given
by $1/ \kappa\bar{n} $, where $\bar{n} = 5  \pm 0.2$ in the main experiment (see Supplement Sec.~II)\tabularnewline
$\kappa^{-1}$ & $44.0 \pm0.06$\,ns & Readout cavity lifetime\tabularnewline
$T_{\mathrm{int}}$ & 260.0\,ns & Integration time of the measurement record, set in the controller at the beginning of the experiment \tabularnewline
$\Gamma_{\mathrm{BG}}^{-1}$ & $0.99\pm0.06\,\mathrm{\mu s}$ & Average time the atom rests in $\ket{\mathrm{G}}$ before an excitation
to $\ket{\mathrm{B}}$, see Fig.~2b\tabularnewline
$\Delta t_{\mathrm{mid}}$ & $3.95\,\mathrm{\mu s}$ & No-click duration for reaching $Z_\mathrm{GD} = 0$ in the flight of the quantum jump from $\ket{\mathrm{G}}$ to $\ket{\mathrm{D}}$, in the full
presence of $\Omega_{\mathrm{DG}}$, see Fig.~3b\tabularnewline
$\Gamma_{\mathrm{GD}}^{-1}$ & $30.8\pm0.4\,\mathrm{\mu s}$ & Average time the atom stays in $\ket{\mathrm{D}}$ before returning
to $\ket{\mathrm{G}}$ and being detected, see Fig.~2b\tabularnewline
$T_{1}^{\mathrm{D}}$ & $116\pm5\,\mathrm{\mu s}$ & Energy relaxation time of $\ket{\mathrm{D}}$\tabularnewline
$T_{2\mathrm{R}}^{\mathrm{D}}$ & $120\pm5\,\mathrm{\mu s}$ & Ramsey coherence time of $\ket{\mathrm{D}}$\tabularnewline
$T_{2\mathrm{E}}^{\mathrm{D}}$ & $162\pm6\,\mathrm{\mu s}$ & Echo coherence time of $\ket{\mathrm{D}}$\tabularnewline
$\Gamma_{\mathrm{DG}}^{-1}$ & $220\pm5\,\mathrm{\mu s}$ & Average time between two consecutive $\ket{\mathrm{G}}$ to $\ket{\mathrm{D}}$
jumps\tabularnewline
\end{tabular}
\par\end{centering}
\caption{
\textbf{Summary of timescales.}
List of the characteristic timescales involved in the catch and reverse experiment. 
The Hamiltonian parameters of the system are summarized in Sec.~I of the Supplementary Information. 
\label{tab:summary-timescales}
}
\end{table}

\clearpage
\nociteMethods{apsrev41Control} 
\bibliographystyleMethods{apsrev4-1}
\bibliographyMethods{mycontrol,biblio}

\end{document}


\renewcommand{\figurename}{\textbf{Figure}}
\renewcommand{\tablename}{\textbf{Table}}

\title{Supplementary Information for:\\ ``To catch and reverse a quantum jump mid-flight''}

\author{Z.K. Minev}
 \email{zlatko.minev@aya.yale.edu | zlatko-minev.com}
 \affiliation{Department of Applied Physics, Yale University, New Haven, Connecticut 06511, USA}
  \author{S.O. Mundhada}
   \affiliation{Department of Applied Physics, Yale University, New Haven, Connecticut 06511, USA}
\author{S. Shankar}
   \affiliation{Department of Applied Physics, Yale University, New Haven, Connecticut 06511, USA}
\author{P. Reinhold}
   \affiliation{Department of Applied Physics, Yale University, New Haven, Connecticut 06511, USA}
\author{R. Guti\'errez-J\'auregui}
   \affiliation{The Dodd-Walls Centre for Photonic and Quantum Technologies, Department of Physics, University of Auckland, Private Bag 92019, Auckland, New Zealand}
\author{R.J. Schoelkopf}
   \affiliation{Department of Applied Physics, Yale University, New Haven, Connecticut 06511, USA}
\author{M. Mirrahimi}
	\affiliation{Yale Quantum Institute, Yale University, New Haven, Connecticut 06520, USA}
	\affiliation{QUANTIC team, INRIA de Paris, 2 Rue Simone Iff, 75012 Paris, France}
\author{H.J. Carmichael}
   \affiliation{The Dodd-Walls Centre for Photonic and Quantum Technologies, Department of Physics, University of Auckland, Private Bag 92019, Auckland, New Zealand}
\author{M.H. Devoret}
   \affiliation{Department of Applied Physics, Yale University, New Haven, Connecticut 06511, USA}

\date{\today}

\maketitle
\vspace{5em}
\tableofcontents
\clearpage

\onecolumngrid
\makeatletter
\makeatletter
\renewcommand{\fnum@figure}{\textbf{Supplementary Figure~\thefigure}}
\renewcommand{\fnum@table}{\textbf{Supplementary Table~\thetable}}
\makeatother

\renewcommand{\thefigure}{S\arabic{figure}}
\renewcommand{\thetable}{S\arabic{table}}
\setcounter{figure}{0}

\section{Experimental characterization of the system}

\begin{figure}[!b]
\begin{centering}
\ifjournal
	\includegraphics[width=140mm]{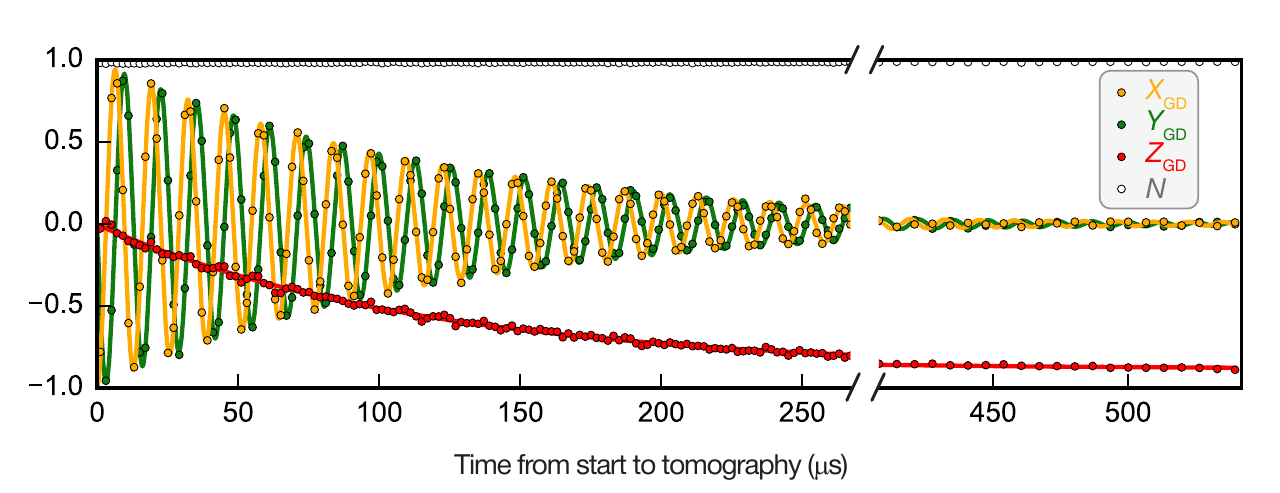}
\else
	\includegraphics[width=140mm]{time-resolved-tomo-D.pdf}
\fi
\caption{\label{fig:time-resolved-tomo-D}
\textbf{Control experiment: time-resolved tomogram of the free evolution of a DG superposition.}
The atom is prepared in $\frac{1}{\sqrt{2}} \left( \D - \G \right)$ and tomography is performed after a varied delay.  Dots: reconstructed conditional GD tomogram ($X_\mathrm{DG},Y_\mathrm{DG},$ and $Z_\mathrm{DG}$) and population in DG manifold,  $N$ (see Methods). Solid lines: theoretical fits.
}
\end{centering}
\end{figure}

\textit{Hamiltonian of the device.}
The two-transmon, single-readout-cavity system is well described, in the low-excitation manifold, by the approximate dispersive Hamiltonian\cite{Nigg2012}${}^{,}$\footnote{Z. K. Minev \textit{et al.}, Energy-participation approach to the design of quantum Josephson circuits, in prep.}: \begin{eqnarray}
\hat{H}/\hbar&=& \omega_{\mathrm{B}}\hat{b}^{\dagger}\hat{b}-\frac{1}{2}\alpha_{\mathrm{B}}\hat{b}^{\dagger2}\hat{b}^{2}+\omega_{\mathrm{D}}\hat{d}^{\dagger}\hat{d}-\frac{1}{2}\alpha_{\mathrm{D}}\hat{d}^{\dagger2}\hat{d}^{2}-\chi_{\mathrm{DB}}\hat{b}^{\dagger}\hat{b}\hat{d}^{\dagger}\hat{d}\label{eq:Hamiltonian-of-sys}\\
 &&+\left(\omega_{\mathrm{C}}+\chi_{\mathrm{B}}\hat{b}^{\dagger}\hat{b}+\chi_{\mathrm{D}}\hat{d}^{\dagger}\hat{d}\right)\hat{c}^{\dagger}\hat{c}\,,\nonumber
\end{eqnarray}
where $\omega_{\mathrm{C}}$, $\omega_{\mathrm{B}}$ and $\omega_{\mathrm{D}}$
are the cavity, bright, and dark qubit transition frequencies, $\hat{c}$,
$\hat{b}$ and $\hat{d}$ are the associated ladder operators, and
$\alpha$ and $\chi$ are the modal anharmonicities and dispersive
shifts, respectively. \footnote{
For related work on  three-level superconducting systems, see Refs.~\onlinecite{Gambetta2011-Purcell, Srinivasan2011, Diniz2013, Dumur2015, Roy2017-3qubits, Zhang2017}.}
The states $\ket{\mathrm{B}}$ and $\ket{\mathrm{D}}$
correspond to one excitation in the bright ($\hat{b}^{\dagger}\ket 0$)
and dark ($\hat{d}^{\dagger}\ket 0$) modes, respectively.
The measured parameters of the device and the mode coherences are summarized in Table~\ref{tab:system-params}.

\begin{table}
\begin{centering}
\renewcommand*\arraystretch{1.5}\begin{tabular}{rl|rl|rl}
\multicolumn{2}{c|}{\textbf{Readout cavity}} & \multicolumn{2}{c|}{\textbf{BG transition}} & \multicolumn{2}{c}{\textbf{DG transition}}\tabularnewline
\hline
\hline
\multicolumn{6}{c}{\textbf{\rule{0pt}{5ex}Mode frequencies and non-linear parameters}}\tabularnewline
\hline
\textbf{\rule{0pt}{4ex}}$\mkern11mu\omega_{\mathrm{C}}/2\pi=$ & $8979.640\text{\,MHz}$ & ~$\omega_{\mathrm{BG}}/2\pi=$ & $5570.349\text{\,MHz}$ & ~$\omega_{\mathrm{DG}}/2\pi=$ & $4845.255\text{\,MHz}$\tabularnewline
 &  & $\chi_{\mathrm{B}}/2\pi=$ & $-5.08\pm0.2\,\mathrm{MHz}$~ & $\chi_{\mathrm{D}}/2\pi=$ & $-0.33\pm0.08\,\mathrm{MHz}$\tabularnewline
 &  & $\alpha_{\mathrm{B}}/2\pi=$ & $195\pm2\,\mathrm{MHz}$ & $\alpha_{\mathrm{D}}/2\pi=$ & $152\pm2\,\mathrm{MHz}$\tabularnewline
 &  & \multicolumn{4}{c}{$\chi_{\mathrm{DB}}/2\pi=61\pm2\,\mathrm{MHz}$}\tabularnewline
\multicolumn{6}{c}{\textbf{\rule{0pt}{5ex}Coherence related parameters}}\tabularnewline
\hline
\textbf{\rule{0pt}{5ex}}$\mkern12mu\kappa/2\pi$= & $3.62\pm0.05\,\mathrm{MHz}\mkern12mu$ & $T_{\mathrm{1}}^{\mathrm{B}}=$ & $28\pm2\,\mathrm{\mu s}$ & $T_{\mathrm{1}}^{\mathrm{D}}=$ & $116\pm5\,\mathrm{\mu s}$\tabularnewline
$\eta=$ & $0.33\pm0.03$ & $T_{\mathrm{2\mathrm{R}}}^{\mathrm{B}}=$ & $18\pm1\,\mathrm{\mu s}$ & $T_{\mathrm{2\mathrm{R}}}^{\mathrm{D}}=$ & $120\pm5\,\mathrm{\mu s}$\tabularnewline
$T_{\mathrm{int}}=$ & $260.0\,\mathrm{ns}$ & $T_{\mathrm{2\mathrm{E}}}^{\mathrm{B}}=$ & $25\pm2\,\mathrm{\mu s}$ & $T_{\mathrm{2\mathrm{E}}}^{\mathrm{D}}=$ & $162\pm6\,\mathrm{\mu s}$\tabularnewline
$n_{\mathrm{th}}^{\mathrm{C}}\le$ & $0.0017\pm0.0002$ & $n_{\mathrm{th}}^{\mathrm{B}}\le$ & $0.01\pm0.005$ & $n_{\mathrm{th}}^{\mathrm{D}}\le$ & $0.05\pm0.01$\tabularnewline
\multicolumn{6}{c}{\textbf{\rule{0pt}{5ex}Drive amplitude and detuning parameters}}\tabularnewline
\hline
\textbf{\rule{0pt}{5ex}}$\mkern12mu\bar{n}=$ & $5.0\pm0.2\mkern12mu$ & ~$\Omega_{\mathrm{B}0}/2\pi=$ & $1.2\pm0.01\,\mathrm{MHz}$ & ~$\Omega_{\mathrm{DG}}/2\pi=$ & $20\pm2\,\mathrm{kHz}$\tabularnewline
 &  & ~$\Omega_{\mathrm{B}1}/2\pi=$ & $600\pm10\,\mathrm{kHz}$ &  & \tabularnewline
$\Delta_{\mathrm{R}}=$ & $\chi_{\mathrm{B}}$ & ~$\Delta_{\mathrm{B}1}/2\pi=$ & $-30.0\,\text{MHz}$ & ~$\Delta_{\mathrm{DG}}/2\pi=$ & $-275.0\,\text{kHz}$\tabularnewline
\end{tabular}
\par\end{centering}
\caption{\textbf{\label{tab:system-params}Compilation of the experimental
parameters.}}
\end{table}

\vspace{10em}

\begin{figure}[hbt]
\begin{centering}
\ifjournal
	\includegraphics[width=88mm]{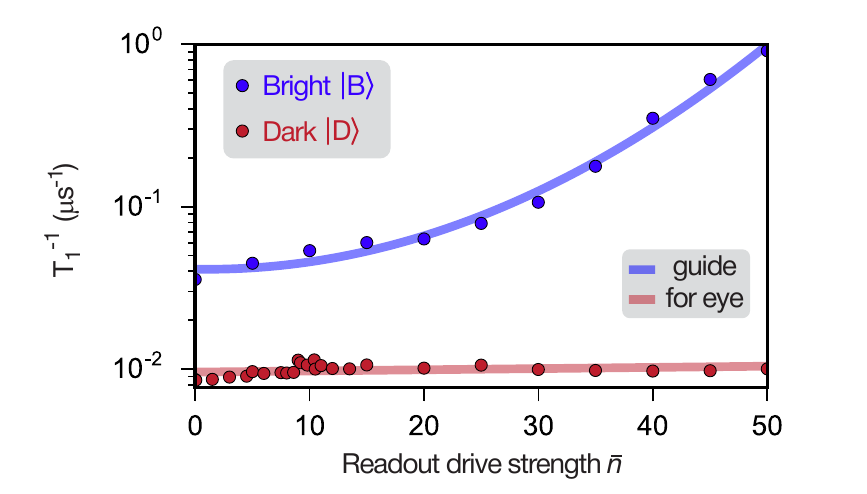}
\else
	\includegraphics[width=88mm]{T1-vs-nbar.pdf}
\fi
\caption{\label{fig:T1-vs-nbar}
\textbf{Measurement-induced energy relaxation $T_1(\bar{n})$.}
Energy relaxation rate ($T_1^{-1}$) of $\B$ (blue dots) and $\D$ (red dots) as a function of $\bar{n}$, measured with the following protocol:
after the atom is prepared in either $\B$ or $\D$, the readout tone ($\mathrm R$) is turned on for duration $t_\mathrm{read}$ with amplitude $\bar{n}$ (corresponding to the number of steady-state photons in the readout cavity when excited on resonance), whereafter the population of the initial state is measured.
As in all other experiments, the readout drive is applied at the $\B$ cavity frequency ($\omega_\mathrm{C}-\chi_\mathrm{B}$).
The relaxation rates are extracted from exponential fits of the population decay as a function of $t_\mathrm{read}$, from $1.3\times10^7$ experimental realizations.
The solids lines are guides to the eye:
blue line indicates the rapid degradation of $T_1^\mathrm{B}$ as a function of the readout strength, while the red line indicates the nearly constants $T_1^\mathrm{D}$ of the protected dark level.
}
\end{centering}
\end{figure}

\textit{DG coherence \& tomography control.}
In Fig.~\ref{fig:time-resolved-tomo-D}, we show the results of a control experiment where we verified the Ramsey coherence ($T_\mathrm{2R}^\mathrm{D}$) and energy relaxation ($T_\mathrm{1}^\mathrm{D}$) times of the DG transition with our tomography method.
Solid lines are fitted theoretical curves for the free evolution of the prepared initial state $\frac{1}{\sqrt{2}} \left( \D - \G \right)$.
The $T_\mathrm{2R}^\mathrm{D} = 119.2\,\mathrm{\mu s}$ value gained from the simultaneous fit of $X_\mathrm{DG}(t)$ and $Y_\mathrm{DG}(t)$  matches the lifetime independently obtained from a standard $T_\mathrm{2R}$ measurement. Similarly, the  value of  $T_\mathrm{1}^\mathrm{D} = 115.4\,\mathrm{\mu s}$ extracted from an exponential fit of  $Z_\mathrm{DG}(t)$ matches the value obtained from a standard $T_\mathrm{1}$ measurement. We note that our tomography procedure is well calibrated and skew-free, as evident in the zero steady-state values of $X_\mathrm{DG}$ and $Y_\mathrm{DG}$. The steady state $Z_\mathrm{DG}$ corresponds to the  thermal population of the dark state $n_{\mathrm{th}}^{\mathrm{D}}$. It has recently been shown that residual thermal populations in cQED systems can be significantly reduced by properly thermalizing the input-output lines.\cite{Yeh2017Atten,Wang2019-cav-atten}

\textit{Measurement-induced energy relaxation $T_1(\bar{n})$.} Figure~\ref{fig:T1-vs-nbar} shows a characterization of the parasitic measurement-induced energy relaxation  of $\B$ and $\D$.
As is the case in standard cQED systems\cite{Slichter2012, Slichter2016-T1vsNbar, Sank2016-T1vsNbar}, the $\B$ level shows strong $T_1$ degradation\cite{Boissonneault2008, Boissonneault2009-Photon-induced-relax,Verney2018,Lescanne2018} as a function of the readout drive strength $\bar{n}$.
However, the lifetime of the dark state ($\D$) is protected, and remains largely unaffected even at large drive strengths ($\bar{n}\approx50$).

\section{Quantum trajectory theory}
\label{sec:quantum-traj-theory}

\subsection{Fluorescence monitored by photon counts}
\label{sec:fluorescence_monitored_by_photon_counts}
\subsubsection{Coherent Bright drive} 
The experiments with trapped ions \cite{Nagourney1986,Sauter1986,Bergquist1986} monitor intermittent fluorescence from the bright state $|{\rm B}\rangle$ to track jumps between $|{\rm G}\rangle$ and $|{\rm D}\rangle$.\cite{Cook1985} In the simplest three-level scheme,\cite{Bergquist1986} and using coherent radiation to excite both  transitions, the master equation for the reduced density operator $\rho$  of the three-level system, written in the interaction picture, is
\begin{equation}
\frac{\mathrm{d}\rho}{\mathrm{d}t}=(i\hbar)^{-1}[\hat H_{\rm drive},\rho]+\gamma_{\rm B}{\cal D}\left[|{\rm G}\rangle\langle{\rm B}|\right]\rho+\gamma_{\rm D}{\cal D}\left[|{\rm G}\rangle\langle{\rm D}|\right]\rho,
\end{equation}
where ${\cal D}[\hat \xi]\mkern3mu\cdot=\hat \xi\cdot \xi^\dagger-\frac12\{\hat \xi^\dagger\hat \xi,\cdot\mkern3mu\}$ denotes the Lindblad superoperator, $\gamma_{\rm B}$ and $\gamma_{\rm D}$ are radiative decay rates, and
\begin{equation}
\hat H_{\rm drive}=i\hbar\frac{\Omega_{\rm BG}}2\big(|{\rm B}\rangle\langle {\rm G}|-|{\rm G}\rangle\langle{\rm B}|\big)+i\hbar\frac{\Omega_{\rm DG}}2\big(|{\rm D}\rangle\langle{\rm G}|-|{\rm G}\rangle\langle{\rm D}|\big),
\label{eqn:drive_coherent_fluorescence}
\end{equation}
with $\Omega_{\rm BG}$ and $\Omega_{\rm DG}$ the Rabi drives. The quantum trajectory description\cite{Carmichael1993,Dalibard1992-original-traj,Dum1992} unravels $\rho$ into an ensemble of pure states whose ket vectors evolve along stochastic paths conditioned on the clicks of imaginary photon detectors that monitor fluorescence from $|{\rm B}\rangle$ (and much less frequently from $|{\rm D}\rangle$). In recognition of each click the ket vector is reset to $|{\rm G}\rangle$, while otherwise it follows a deterministic evolution as a coherent superposition,
\begin{equation}
|\psi(\Delta t_{\rm catch})\rangle=C_{\rm G}(\Delta t_{\rm catch})|{\rm G}\rangle+C_{\rm B}(\Delta t_{\rm catch})|{\rm B}\rangle+C_{\rm D}(\Delta t_{\rm catch})|{\rm D}\rangle,
\label{eqn:GBD-ket}
\end{equation}
where $C_{\rm G}(0)=1$, $C_{\rm B}(0)=C_{\rm D}(0)=0$, with $\Delta t_{\rm catch}=0$ marking the time of the last click reset, and
\begin{equation}
i\hbar\frac{\mathrm{d}|\psi\rangle}{\mathrm{d}\Delta t_{\rm catch}}=\left(\mkern-2mu \hat H_{\rm drive}-i\hbar\frac{\gamma_{\rm B}}2|{\rm B}\rangle\langle{\rm B}|-i\hbar\frac{\gamma_{\rm D}}2|{\rm D}\rangle\langle{\rm D}|\right)\mkern-3mu|\psi\rangle.
\label{eqn:between_click_equation}
\end{equation}
The norm of the ket $|\psi(\Delta t_{\rm catch})\rangle$ is not preserved, but rather gives the probability that the evolution will continue, with no interruption by further clicks, up to time $\Delta t_{\rm catch}$; clearly it must decay with this probability to zero.
If we then define
\begin{equation}
W_{\rm DG}(\Delta t_{\rm catch})\equiv\frac{C_{\rm D}(\Delta t_{\rm catch})}{C_{\rm G}(\Delta t_{\rm catch})},
\label{eqn:W_DG_definition}
\end{equation}
the \textit{normalized} ket vector in the GD-subspace has Bloch vector components
\begin{eqnarray}
{\rm Z}_{\rm GD}(\Delta t_{\rm catch})&=&\frac{W_{\rm DG}(\Delta t_{\rm catch})-W_{\rm DG}^{-1}(\Delta t_{\rm catch})}{W_{\rm DG}(\Delta t_{\rm catch})+W_{\rm DG}^{-1}(\Delta t_{\rm catch})},
\label{eqn:Z}\\
\noalign{\vskip4pt}
{\rm X}_{\rm GD}(\Delta t_{\rm catch})&=&\frac2{W_{\rm DG}(\Delta t_{\rm catch})+W_{\rm DG}^{-1}(\Delta t_{\rm catch})},
\label{eqn:X}\\
\noalign{\vskip6pt}
{\rm Y}_{\rm GD}(\Delta t_{\rm catch})&=&0,
\label{eqn:Y}
\end{eqnarray}
where, using Eqs.~(\ref{eqn:GBD-ket}) and (\ref{eqn:between_click_equation}),
\begin{equation}
\frac{d}{d\Delta t_{\rm catch}}\mkern-2mu\left(
\begin{matrix}
C_{\rm G}\\
C_{\rm B}\\
C_{\rm D}
\end{matrix}
\right)=\frac12\left(
\begin{matrix}
0&-\Omega_{\rm BG}&-\Omega_{\rm DG}\\
\Omega_{\rm BG}&-\gamma_{\rm B}&0\\
\Omega_{\rm DG}&0&-\gamma_{\rm D}\\
\end{matrix}
\right)\mkern-4mu\left(
\begin{matrix}
C_{\rm G}\\
C_{\rm B}\\
C_{\rm D}
\end{matrix}
\right).
\label{eqn:3x3_system}
\end{equation}
In general this $3\times3$ system does not have a closed solution in simple form, although there is a particularly simple solution under conditions  that produce intermittent fluorescence, i.e., rare jumps from $|{\rm G}\rangle$ to $|{\rm D}\rangle$ (``shelving'' in the dark state \cite{Nagourney1986}) interspersed as intervals of fluorescence ``off'' in a background of fluorescence ``on''. The conditions follow naturally if $|{\rm D}\rangle$ is a metastable state,\cite{Nagourney1986,Sauter1986,Bergquist1986,Macieszczak2016} whose lifetime $\gamma_{\rm D}^{-1}$ is extremely long on the scale of the mean time  between photon detector clicks for a weak $\Omega_{\rm BG}$ Rabi drive,
\begin{equation}
\Gamma_\mathrm{BG}^{-1} = \left(\frac{\Omega_{\rm BG}^2}{\gamma_{\rm B}}\right)^{-1}\;. 
\label{eq:GammaBG}
\end{equation}
Thus, for $(\Omega_{\rm DG},\gamma_{\rm D})\ll\Omega_{\rm BG}^2/\gamma_{\rm B}\ll\gamma_{\rm B}$, Eq.~(\ref{eqn:3x3_system}) yields the equation of motion
\begin{equation}
\frac{\mathrm{d}W_{\rm DG}}{\mathrm{d}\Delta t_{\rm catch}}=\frac{\Omega_{\rm BG}^2}{2\gamma_{\rm B}}W_{\rm DG}+\frac{\Omega_{\rm DG}}2,
\label{eqn:W_DG_equation_coherent}
\end{equation}
whose solution for the click reset initial condition, $W_{\rm DG}(0)=0$, is
\begin{equation}
W_{\rm DG}(\Delta t_{\rm catch})=\frac{\Omega_{\rm DG}}{\Omega^2_{\rm BG}/\gamma_{\rm B}}\left[\exp\left(\frac{\Omega^2_{\rm BG}}{2\gamma_{\rm B}}\Delta t_{\rm catch}\right)-1\right],
\label{eqn:W_DG_coherent}
\end{equation}
from which a long enough interval with no clicks gives $W_{\rm DG}(\Delta t_{\rm catch})\gg1$ and leads to the conclusion that the ket vector is $|{\rm D}\rangle$.
The time scale for the transition, $\Delta t_{\rm mid}$, is defined by $Z_\mathrm{GD}(\Delta t_{\rm mid})=0$, which corresponds to $W_{\rm DG}(\Delta t_{\rm mid})=1$.\cite{Porrati1987} 
Simply inverting Eq.~\eqref{eqn:W_DG_coherent} produces the formula
\begin{equation}
\Delta t_{\rm mid}=\left(\frac{\Omega^2_{\rm BG}}{2\gamma_{\rm B}}\right)^{-1}\ln\left(\frac{\Omega^2_{\rm BG}/\gamma_{\rm B}}{\Omega_{\rm DG}}+1\right),
\label{eqn:t_mid_coherent}
\end{equation}
but strong monitoring, $\Omega^2_{\rm BG}/\gamma_{\rm B}\gg\Omega_{\rm DG}$, allows the $-1$, in Eq.~(\ref{eqn:W_DG_coherent}), and $+1$, in Eq.~(\ref{eqn:t_mid_coherent}), to be dropped. Equations (\ref{eqn:Z})--(\ref{eqn:Y}), (\ref{eqn:W_DG_coherent}), and (\ref{eqn:t_mid_coherent}) then provide simple formulas for the continuous, deterministic, and coherent evolution of the quantum jump when completed:
\begin{eqnarray}
{\rm Z}_{\rm GD}(\Delta t_{\rm catch})&=&{\rm tanh}\left[\frac{\Omega^2_{\rm BG}}{2\gamma_B}(\Delta t_{\rm catch}-\Delta t_{\rm mid})\right],
\label{eqn:Z_approx}\\
\noalign{\vskip4pt}
{\rm X}_{\rm GD}(\Delta t_{\rm catch})&=&{\rm sech}\left[\frac{\Omega^2_{\rm BG}}{2\gamma_B}(\Delta t_{\rm catch}-\Delta t_{\rm mid})\right],
\label{eqn:X_approx}\\
\noalign{\vskip6pt}
{\rm Y}_{\rm GD}(\Delta t_{\rm catch})&=&0.
\label{eqn:Y_approx}
\end{eqnarray}
These formulas execute a perfect jump, ${\rm Z}_{\rm GD}(\infty)=1$, ${\rm X}_{\rm GD}(\infty)={\rm Y}_{\rm GD}(\infty)=0$. The ideal arises from the assumed strong monitoring, $\Omega_{\rm DG}\ll\Omega^2_{\rm BG}/\gamma_{\rm B}$. Departures from it can be transparently analyzed by adopting an incoherent Bright drive, see Sec.~\ref{sec:incoherent-B-drive}. 
An elegant analysis of the no-click evolution for arbitrary amplitude of the Dark Rabi drive can be found in Refs.~\onlinecite{Ruskov2007} and~\onlinecite{Ruskov2009-unpublished}. For an interesting connection to of the three-level intermittent dynamics to dynamical phase transitions, see Refs.~\onlinecite{Lesanovsky2013} and~\onlinecite{Garrahan2018}.

\textit{Application of the photon counting model to the experiment.} 
The photon-counting theory presented in this section  provides the background to the experiment along with a link to the original ion experiments.
It captures a core set of the ideas, even though the monitoring of $|\rm B\rangle$ implemented in the experiment is diffusive --- the opposite limit of the point-process description presented here, see Sec.~\ref{sec:Cavity-Monitor}. 
Nevertheless, the photon-counting theory even provides a quantitative first approximation of the experimental results. 
For definitiveness, consider the flight of the quantum jump shown in Fig.~3b.
The measured mid-flight time, $\Delta t_\mathrm{mid} = 3.95~\mathrm{\mu s}$, is predicted, in a first approximation, by Eq.~\eqref{eqn:t_mid_coherent}.
Using the (independently measured) values of the experimental parameters, summarized in Table~\ref{tab:system-params} (setting $\Omega_\mathrm{BG}$ equal to  $\Omega_\mathrm{B0} = 2\pi\times1.2$~MHz, the BG drive when the atom is not in $\ket{\rm B}$) and extracting the effective measurement rate of $|\rm B\rangle$, $\gamma_\mathrm{B} = 2\pi\times 9.0$~MHz (which follows from Eq.~\eqref{eq:GammaBG} where $\Gamma_\mathrm{BG} = 2\pi\times1.01$~MHz, the average click rate on the BG transition), Eq.~\eqref{eqn:t_mid_coherent} predicts $\Delta t_{\rm mid} \approx 4.3~\mathrm{\mu s}$ --- in fair agreement with the observed value $\Delta t_{\rm mid}  = 3.95~\mathrm{\mu s}$. 
The photodetection theory presented in in Sec.~\ref{sec:incoherent-B-drive} further improves the agreement. These calculations serve to generally illustrate the theory and ideas of the experiment; the quantitative comparison between theory and experiment is only presented in Sec.~\ref{sec:Comparison-theory-exp}.

\subsubsection{Incoherent Bright drive} 
\label{sec:incoherent-B-drive}
If the coherent Rabi drive $\Omega_{\rm BG}$ is replaced by an incoherent drive $\Gamma_{\rm BG}$, the master equation in the interaction picture becomes
\begin{equation}
\frac{\mathrm{d}\rho}{\mathrm{d}t}=(i\hbar)^{-1}[\hat H_{\rm drive},\rho]+\Gamma_{\rm BG}{\cal D}[|{\rm B}\rangle\langle{\rm G}|]\rho+(\gamma_{\rm B}+\Gamma_{\rm BG}){\cal D}[|{\rm G}\rangle\langle{\rm B}|]\rho+\gamma_{\rm D}{\cal D}[|{\rm G}\rangle\langle{\rm D}|]\rho,
\end{equation}
where
\begin{equation}
\hat H_{\rm drive}=i\hbar\frac{\Omega_{\rm DG}}2\big(|{\rm D}\rangle\langle{\rm G}|-|{\rm G}\rangle\langle{\rm D}|\big).
\end{equation}
The previous weak drive assumption, $\Omega_{\rm BG}^2/\gamma_{\rm B}\ll\gamma_{\rm B}$, is now carried over with the assumption $\Gamma_{\rm BG}\ll\gamma_{\rm B}$, which says that the time between clicks in fluorescence is essentially the same as the time separating photon absorptions from the incoherent drive, as absorption is rapidly followed by fluorescence ($\gamma_{\rm B}+\Gamma_{\rm BG}\gg\Gamma_{\rm BG}$). This brings a useful simplification, since, following each reset to $|{\rm G}\rangle$, the unnormalized ket evolves in the GD-subspace,
\begin{equation}
i\hbar\frac{\mathrm{d}|\psi\rangle}{\mathrm{d}\Delta t_{\rm catch}}=\left(\mkern-2mu H_{\rm drive}-i\hbar\frac{\Gamma_{\rm BG}}2|{\rm G}\rangle\langle{\rm G}|-i\hbar\frac{\gamma_{\rm D}}2|{\rm D}\rangle\langle{\rm D}|\right)\mkern-3mu|\psi\rangle,
\end{equation}
thus replacing Eqs.~(\ref{eqn:3x3_system}) and (\ref{eqn:W_DG_equation_coherent}) by the simpler $2\times2$ system
\begin{equation}
\frac{\mathrm{d}}{d\Delta t_{\rm catch}}\mkern-2mu\left(
\begin{matrix}
C_{\rm G}\\
C_{\rm D}
\end{matrix}
\right)=\frac12\left(
\begin{matrix}
-\Gamma_{\rm BG}&-\Omega_{\rm DG}\\

\Omega_{\rm DG}&-\gamma_{\rm D}\\
\end{matrix}
\right)\mkern-4mu\left(
\begin{matrix}
C_{\rm G}\\
C_{\rm D}
\end{matrix}
\right),
\end{equation}
and, if $\gamma_{\rm D}\ll\Gamma_{\rm BG}$, the equation of motion
\begin{equation}
\frac{dW_{\rm DG}}{\mathrm{d}\Delta t_{\rm catch}}=\frac{\Gamma_{\rm BG}}2W_{\rm DG}+\frac{\Omega_{\rm DG}}2(1+W_{\rm DG}^2),
\label{eqn:W_DG_equation_incoherent}
\end{equation}
with solution, for $W_{\rm DG}(0)=0$,
\begin{equation}
W_{\rm DG}(\Delta t_{\rm catch})=\frac{\exp\left[(V-V^{-1})\Omega_{\rm DG}\Delta t_{\rm catch}/2\right]-1}{V-V^{-1}\exp\left[(V-V^{-1})\Omega_{\rm DG}\Delta t_{\rm catch}/2\right]},
\label{eqn:W_DG_incoherent}
\end{equation}
where
\begin{equation}
V=\frac12\frac{\Gamma_{\rm BG}}{\Omega_{\rm DG}}+\sqrt{\frac14\left(\frac{\Gamma_{\rm BG}}{\Omega_{\rm DG}}\right)^2-1}.
\end{equation}
In Ref.~\onlinecite{Ruskov2007}, a general form of the Bloch vector equations for arbitrary amplitude of the Rabi drive was found.
Inversion of the condition $W_{\rm DG}(\Delta  t_{\rm mid})=1$ gives the characteristic time scale
\begin{equation}
\Delta t_{\rm mid}=2\left[(V-V^{-1})\Omega_{\rm DG}\right]^{-1}\ln\left(\frac{V+1}{V^{-1}+1}\right).
\label{eqn:t_mid_incoherent}
\end{equation}
Equations (\ref{eqn:W_DG_incoherent})--(\ref{eqn:t_mid_incoherent}) replace Eqs.~(\ref{eqn:W_DG_coherent}) and (\ref{eqn:t_mid_coherent}); although, under strong monitoring ($\Gamma_{\rm BG}\gg\Omega_{\rm DG}$), they revert to these results with the substitution $\Omega_{\rm BG}^2/2\gamma_{\rm B}\to\Gamma_{\rm BG}/2$, recovering Eqs.~(\ref{eqn:Z})--(\ref{eqn:Y}) with the same substitution. More generally, $W_{\rm DG}(\Delta t_{\rm catch})$ goes to infinity at finite $\Delta t_{\rm catch}$, changes sign, and returns  from infinity to settle on the steady value $W_{\rm DG}(\infty)=-V$. The singular behavior marks a trajectory passing through the north pole of Bloch sphere. It yields the long-time solution
\begin{eqnarray}
{\rm Z}_{\rm GD}(\infty)=\sqrt{1-4\left(\frac{\Omega_{\rm DG}}{\Gamma_{\rm BG}}\right)^2},\qquad
{\rm X}_{\rm GD}(\infty)=-2\frac{\Omega_{\rm DG}}{\Gamma_{\rm BG}},\qquad
{\rm Y}_{\rm GD}(\infty)=0,
\label{eq:XYZ_GD-inf}
\end{eqnarray}
in contrast to the perfect jump of Eqs.~(\ref{eqn:Z_approx})--(\ref{eqn:Y_approx}).

\subsubsection{Dark drive off}
Turing the Dark drive off shortly after a click reset demonstrates the connection between the flight of a quantum jump and a projective measurement. From the point of view of the trajectory equations, the only change is the setting of $\Omega_{\rm DG}$ to zero at time $\Delta t_{\rm on}$ on the right-hand side of Eqs.~(\ref{eqn:W_DG_equation_coherent}) and (\ref{eqn:W_DG_equation_incoherent}). Subsequently, $W_{\rm DG}(\Delta t_{\rm catch})$ continues its exponential growth at rate $\Omega_{\rm BG}^2/2\gamma_B$ [Eq.~(\ref{eqn:W_DG_equation_coherent})] or $\Gamma_{\rm BG}/2$ [Eq.~(\ref{eqn:W_DG_equation_incoherent})]. Equations (\ref{eqn:Z})--(\ref{eqn:Y})   for the GD Bloch components still hold, 
but now with
\begin{equation}
\Delta t_{\rm mid}=\left(\frac{\Omega_{\rm BG}^2}{2\gamma_B},\frac{\Gamma_{\rm BG}}2\right)^{-1}\ln\big[W_{\rm DG}^{ -2}(\Delta t_{\rm on})\big]\,,
\end{equation}
which can provide an estimate of $\Delta t'_\mathrm{mid}$, specifying the time at which $Z_\mathrm{GD} = 0$.

The evolution during $\Delta t_\mathrm{off}$, in the absence of $\Omega_\mathrm{DG}$, in effect realizes a projective measurement of whether the state of the atom is $|{\rm G}\rangle$ or $|{\rm D}\rangle$, where the normalized state at $\Delta t_{\rm on}$ is
\begin{equation}
\frac{|\psi(\Delta t_{\rm on})\rangle}{\sqrt{{\cal N}(\Delta t_{\rm on})}}=\frac{C_{\rm G}(\Delta t_{\rm on})|{\rm G}\rangle+C_{\rm D}(\Delta t_{\rm on})|{\rm D}\rangle}{\sqrt{{\cal N}(\Delta t_{\rm on})}},
\label{eqn:initial_state}
\end{equation}
with ${\cal N}(\Delta t_{\rm on})=C_{\rm G}^2(\Delta t_{\rm on})+C_{\rm D}^2(\Delta t_{\rm on})$ the probability for the jump to reach $\Delta t_{\rm catch}=\Delta t_{\rm on}$ after a click reset to $|{\rm G}\rangle$ at $\Delta t_{\rm catch}=0$. The probability for the jump to continue to $\Delta t_{\rm catch}>\Delta t_{\rm on}$ (given $\Delta t_{\rm on}$ is reached) is then
\begin{equation}
\label{eq:completion-prob}
\frac{{\cal N}(\Delta t_{\rm catch})}{{\cal N}(\Delta t_{\rm on})}=\frac{C_{\rm D}^2(\Delta t_{\rm on})}{{\cal N}(\Delta t_{\rm on})}+\frac{C_{\rm G}^2(\Delta t_{\rm on})}{{\cal N}(\Delta t_{\rm on})}\exp\mkern-2mu\left[-\left(\frac{\Omega_{\rm BG}^2}{\gamma_{\rm B}},\Gamma_{\rm BG}\right)\mkern-2mu\Delta t_{\rm catch}\right].
\end{equation}

\subsubsection{Completed and aborted evolutions of the jump transition}
 In this simple model, the probability for the trajectory to complete --- for the measurement to yield the result $|{\rm D}\rangle$ --- is obtained in the limit $\Delta t_\mathrm{catch}\rightarrow \infty$, 
 and, as expected,  is equal to the probability to occupy the state $|{\rm D}\rangle$ at time $\Delta t_{\rm on}$; i.e., the completion probability is $P_\mathrm{D}(\Delta t_{\rm on}) = C_{\rm D}^2(\Delta t_{\rm on})/{{\rm Norm}(\Delta t_{\rm on})}$.
It is helpful to illustrate this idea with an example. Consider the catch experiment of Fig.~3b in the absence of the Dark Rabi drive, $\Omega_\mathrm{DG}$. From $Z_\mathrm{GD} $, we can  estimate that out of all the trajectories that pass though the $\Delta t_{\rm on}$ mark approximately $P_\mathrm{D}(\Delta t_{\rm on})= (1+Z_\mathrm{GD}(\Delta t_{\rm on}))/2 \approx 8\%$ fully complete without an interruption. On the other hand, for those that pass the $\Delta t'_\mathrm{mid}$ mark, approximately 50\%  complete. It follows from Eq.~\eqref{eq:completion-prob}, that the  probability of the evolution to complete increases the further along the trajectory is. 
Although some of the jump evolutions abort at random, importantly, every single jump evolution that completes, and is thus recorded as a jump,  follows \textit{not} a random but an identical path in Hilbert space, i.e., a deterministic one. This path (of \textit{any}  jump) is determined   by Eq.~\eqref{eqn:W_DG_incoherent}, or, in the simpler model, by  the Eqs.~\eqref{eqn:Z_approx}-\eqref{eqn:Y_approx} for the components of the GD Bloch vector.

\subsection{Bright state monitored by dispersive cavity readout}
\label{sec:Cavity-Monitor}
\subsubsection{Stochastic Schr\"odinger equation}
Monitoring the quantum jump through fluorescence photon counts provides a clean and simple way of analyzing the deterministic character of the evolution. It is prohibitively challenging for an experiment, though, as the time origin $\Delta t_{\rm catch}=0$ is set by the click reset to $|{\rm G}\rangle$, and in an ensemble of measurements, all resets must be aligned on the very last click before an interval of deterministic evolution (Eqs.~(\ref{eqn:Z})--(\ref{eqn:Y})) in order for $\Delta t_{\rm mid}$ to be aligned over the ensemble; low detection efficiency---on the order of $10^{-3}$ or less\cite{Nagourney1986,Sauter1986,Bergquist1986}---in the first ion experiments does not permit this. Monitoring through a dispersive cavity readout provides a robust way of aligning $\Delta t_{\rm catch}=0$, hence $\Delta t_{\rm mid}$, over an ensemble of measurements. Leaving aside imperfections (see below), the master equation in the interaction picture is
\begin{equation}
\frac{\mathrm{d}\rho}{\mathrm{d}t}=(i\hbar)^{-1}[\hat H_{\rm drive},\rho]+(i\hbar)^{-1}[\hat H_{\rm R},\rho]+\kappa{\cal D}[\hat c]\rho,
\end{equation}
with
\begin{equation}
\hat H_{\rm drive}=i\hbar\left[\frac{\Omega_{\rm BG}(t)}2|{\rm B}\rangle\langle {\rm G}|-\frac{\Omega_{\rm BG}^*(t)}2|{\rm G}\rangle\langle{\rm B}|\right]+i\hbar\frac{\Omega_{\rm DG}}2\big(|{\rm D}\rangle\langle{\rm G}|-|{\rm G}\rangle\langle{\rm D}|\big),
\end{equation}
and
\begin{equation}
\hat H_{\rm R}=-\hbar\Delta_{\rm R}\hat c^\dagger\hat c+i\hbar\frac{\kappa}2\sqrt{\bar n}(\hat c^\dagger-\hat c)+\hbar\big(\chi_{\rm B}|{\rm B}\rangle\langle{\rm B}|+\chi_{\rm D}|{\rm D}\rangle\langle{\rm D}|\big)\hat c^\dagger\hat c,
\end{equation}
where the bi-chromatic drive $\Omega_{\rm BG}(t)=\Omega_{{\rm B}0}+\Omega_{{\rm B}1}\exp(-i\Delta_{{\rm B}1}t)$ replaces the Rabi drive $\Omega_{\rm BG}$ of Eq.~(\ref{eqn:drive_coherent_fluorescence}), $\bar n$ is the mean photon number in the readout cavity when driven on resonance, and $\Delta_{\rm R}$ is the detuning of the probe from the unshifted cavity resonance; the bi-chromatic drive facilitates transitions in both directions between $|{\rm G}\rangle$ and $|{\rm B}\rangle$, given that the bright level shifts when the cavity fills with photons. The quantum  trajectory unraveling monitors the reflected probe with efficiency $\eta$ and accounts for residual photon loss through random jumps; thus, the stochastic Schr\"odinger equation combines a continuous evolution (heterodyne readout channel),
\begin{equation}
\mathrm{d}|\psi\rangle=\left[\frac1{i\hbar}\left(\hat H_{\rm drive}+\hat H_{\rm R}-i\hbar\frac\kappa2\hat c^\dagger\hat c\right)\mkern-3mu\mathrm{d}t+\sqrt\eta\sqrt{\kappa}\mathrm{d}\zeta^{*}\hat c\right]\mkern-3mu|\psi\rangle,
\label{eqn:SSE_continuous}
\end{equation}
where
\begin{equation}
\label{eq:heterodyne-current}
\mathrm{d}\zeta=\sqrt\eta\sqrt\kappa\frac{\langle\psi|\hat a |\psi\rangle}{\langle\psi|\psi\rangle}\mathrm{d}t+\mathrm{d}Z,
\end{equation}
$\mathrm{d}Z$  is a complex Wiener increment, obeying $\operatorname{E}\left[\mathrm{d}Z\right]=0$, $\operatorname{E}\left[\mathrm{d}Z^{2}\right]=0$, and~$\operatorname{Var}\left[\mathrm{d}Z\right]=\operatorname{E}\left[\mathrm{d}Z^{*}\mathrm{d}Z\right]=\mathrm{d}t$, with random jumps (photon loss),
\begin{equation}
|\psi\rangle\to\hat c|\psi\rangle\qquad\hbox{at rate}\qquad(1-\eta)\kappa\frac{\langle\psi|\hat c^\dagger\hat c|\psi\rangle}{\langle\psi|\psi\rangle}.
\end{equation}
The monitored output $\mathrm{d}\zeta$ is scaled---to units of (readout cavity photon number)${}^{1/2}$---and filtered to generate simulated quadratures $I_{\rm rec}$ and $Q_{\rm rec}$ of the measurement record:
\begin{eqnarray}
\label{eq:dI_rec}
\mathrm{d}I_{\rm rec}&=&-\frac{\kappa_{\rm filter}}2\left[I_{\rm rec}\mathrm{d}t-\left(\eta\frac\kappa2\right)^{-1/2}{\rm Re}(\mathrm{d}\zeta)\right],
\label{eqn:simulated_I_int}\\
\label{eq:dQ_rec}
\mathrm{d}Q_{\rm rec}&=&-\frac{\kappa_{\rm filter}}2\left[Q_{\rm rec}\mathrm{d}t-\left(\eta\frac\kappa2\right)^{-1/2}{\rm Im}(\mathrm{d}\zeta)\right],
\label{eqn:simulated_Q_int}
\end{eqnarray}
where $\kappa_{\rm filter}$ is the bandwidth of the experimental readout amplifier chain.

\subsubsection{Independently measured imperfections}
The stochastic Schr\"odinger equation is supplemented by spontaneous and thermal jumps on both the $|{\rm G}\rangle$ to $|{\rm B}\rangle$ and $|{\rm G}\rangle$ to $|{\rm D}\rangle$ transitions, and by pure dephasing of the ${\rm GB}$ and ${\rm GD}$ coherences. With these processes included, the term
\begin{equation}
-i\hbar\left\{\left[\frac{\gamma_{\rm B}}2(n_{\rm th}^{\rm B}+1)+\gamma_{\rm B}^\phi\right]\mkern-3mu|{\rm B}\rangle\langle{\rm B}|+\left[\frac{\gamma_{\rm D}}2(n_{\rm th}^{\rm D}+1)+\gamma_{\rm D}^\phi\right]\mkern-3mu|{\rm D}\rangle\langle{\rm D}|+\frac{\gamma_{\rm B}n_{\rm th}^{\rm B}+\gamma_{\rm D}n_{\rm th}^{\rm D}}2|{\rm G}\rangle\langle{\rm G}|\right\}\nonumber
\end{equation}
is added to the non-Hermitian Hamiltonian $\hat H_{\rm drive}+\hat H_{\rm R}-i\hbar(\kappa/2)\hat c^\dagger\hat c$ on the right-hand side of Eq.~(\ref{eqn:SSE_continuous}), and there are additional random jumps
\begin{eqnarray}
|\psi\rangle\to|{\rm G}\rangle&&\qquad\hbox{at rate}\qquad\gamma_{\rm B}(n_{\rm th}^{\rm B}+1)\frac{\langle\psi|{\rm B}\rangle\langle{\rm B}|\psi\rangle}{\langle\psi|\psi\rangle}+\gamma_{\rm D}(n_{\rm th}^{\rm D}+1)\frac{\langle\psi|{\rm D}\rangle\langle{\rm D}|\psi\rangle}{\langle\psi|\psi\rangle},\\
\noalign{\vskip4pt}
|\psi\rangle\to|{\rm B}\rangle&&\qquad\hbox{at rate}\qquad\gamma_{\rm B}n_{\rm th}^{\rm B}\frac{\langle\psi|{\rm G}\rangle\langle{\rm G}|\psi\rangle}{\langle\psi|\psi\rangle}+2\gamma_{\rm B}^{\phi}\frac{\langle\psi|{\rm B}\rangle\langle{\rm B}|\psi\rangle}{\langle\psi|\psi\rangle},\\
\noalign{\vskip4pt}
|\psi\rangle\to|{\rm D}\rangle&&\qquad\hbox{at rate}\qquad\gamma_{\rm D}n_{\rm th}^{\rm D}\frac{\langle\psi|{\rm G}\rangle\langle{\rm G}|\psi\rangle}{\langle\psi|\psi\rangle}+2\gamma_{\rm D}^{\phi}\frac{\langle\psi|{\rm D}\rangle\langle{\rm D}|\psi\rangle}{\langle\psi|\psi\rangle}.
\end{eqnarray}
The parameters $\gamma_{\rm B,D}$, $n_{\rm th}^{\rm B,D}$, and $\gamma_{\rm B,D}^\phi$ are mapped to the independently measured parameters $T_{\rm B,D}^1$, $n_{\rm th}^{\rm G,D}$, and $T_{2{\rm R}}^{\rm B,D}$ listed in Table \ref{fig:T1-vs-nbar} (see below).

\subsubsection{Leakage from the {\rm GBD}-manifold}
The three-state manifold, $|{\rm G}\rangle$, $|{\rm B}\rangle$, and $|{\rm D}\rangle$, is not strictly closed. Rare transitions to higher excited states of the two-transmon system may occur. This possibility is included with the addition of the further term
\begin{equation}
-i\hbar\left\{\frac{\gamma_{\rm FG}}2|{\rm G}\rangle\langle{\rm G}|+\frac{\gamma_{\rm FD}}2|{\rm D}\rangle\langle{\rm D}|+\frac{\gamma_{\rm GF}+\gamma_{\rm DF}}2|{\rm F}\rangle\langle{\rm F}|\right\}\nonumber
\end{equation}
to the non-Hermitian Hamiltonian, and the associated additional random jumps,
\begin{eqnarray}
|\psi\rangle\to|{\rm F}\rangle&&\qquad\hbox{at rate}\qquad\gamma_{\rm FG}\frac{\langle\psi|{\rm G}\rangle\langle{\rm G}|\psi\rangle}{\langle\psi|\psi\rangle}+\gamma_{\rm FD}\frac{\langle\psi|{\rm D}\rangle\langle{\rm D}|\psi\rangle}{\langle\psi|\psi\rangle},\\
\noalign{\vskip4pt}
|\psi\rangle\to|{\rm G}\rangle&&\qquad\hbox{at rate}\qquad\gamma_{\rm GF}\frac{\langle\psi|{\rm F}\rangle\langle{\rm F}|\psi\rangle}{\langle\psi|\psi\rangle},\\
\noalign{\vskip4pt}
|\psi\rangle\to|{\rm D}\rangle&&\qquad\hbox{at rate}\qquad\gamma_{\rm DF}\frac{\langle\psi|{\rm F}\rangle\langle{\rm F}|\psi\rangle}{\langle\psi|\psi\rangle},
\label{eqn:jumps_to_F}
\end{eqnarray}
where $|{\rm F}\rangle$ is a single catch-all higher excited state.

\section{Comparison between experiment and theory}
\label{sec:Comparison-theory-exp}

\subsection{Simulated data sets}
\textit{Independently measured parameters.}
The parameters used in the simulations are listed in Table~\ref{table:table2}. In most cases they are set to the value at the center of the range quoted in Table \ref{tab:system-params}, but with three exceptions: (i) $T_1^{\rm B}$ and $T_1^{\rm D}$ are set to lower values in response to the photon number dependence of the readout displayed in Fig.~\ref{fig:T1-vs-nbar}; (ii) $\Omega_{\rm DG}/2\pi$ is set higher, but still falls inside the experimental error bars, and (iii) $n_{\rm th}^{\rm C}=0$. Of the three exceptions, only $\Omega_{\rm DG}/2\pi$ has a noticeable effect on the comparison between simulated and experimental data sets.

\textit{Leakage from the {\rm GBD}-manifold.} Additional random jumps to state $|{\rm F}\rangle$ are governed by four parameters that are not independently measured; they serve as fitting parameters, required to bring the simulation into agreement with the asymptotic behavior of ${\rm Z}(\Delta t_{\rm catch})$, which, without leakage to $|{\rm F}\rangle$, settles  to a value higher than is measured in the experiment. The evolution of the ${\rm X}(\Delta t_{\rm catch})$ is largely unaffected by the assignment of these parameters, where any change that does occur can be offset by adjusting $\Omega_{\rm DG}/2\pi$ while staying within the experimental error bars.

\begin{figure}[t]
\begin{centering}
\vskip0.5cm
\includegraphics[width=11cm]{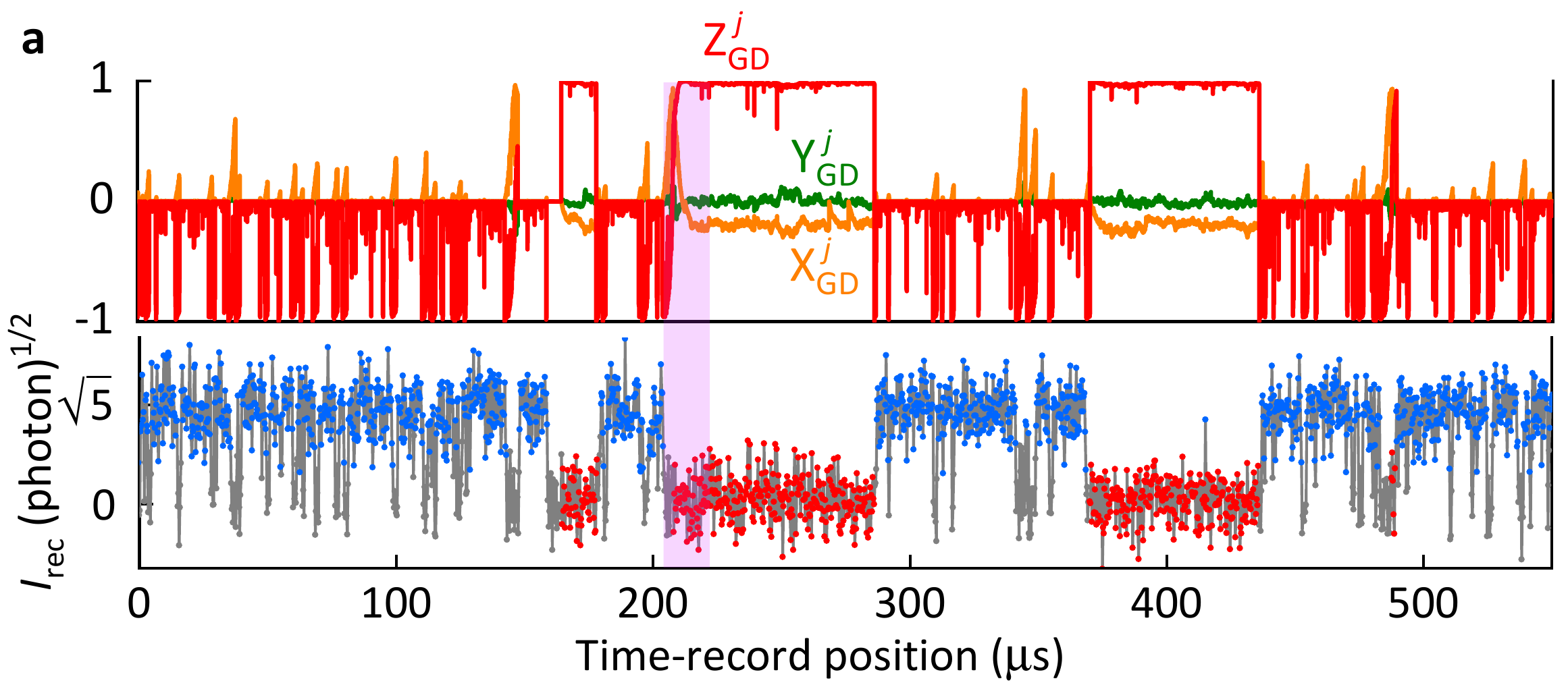}\raise0.0cm\hbox{\includegraphics[width=7cm]{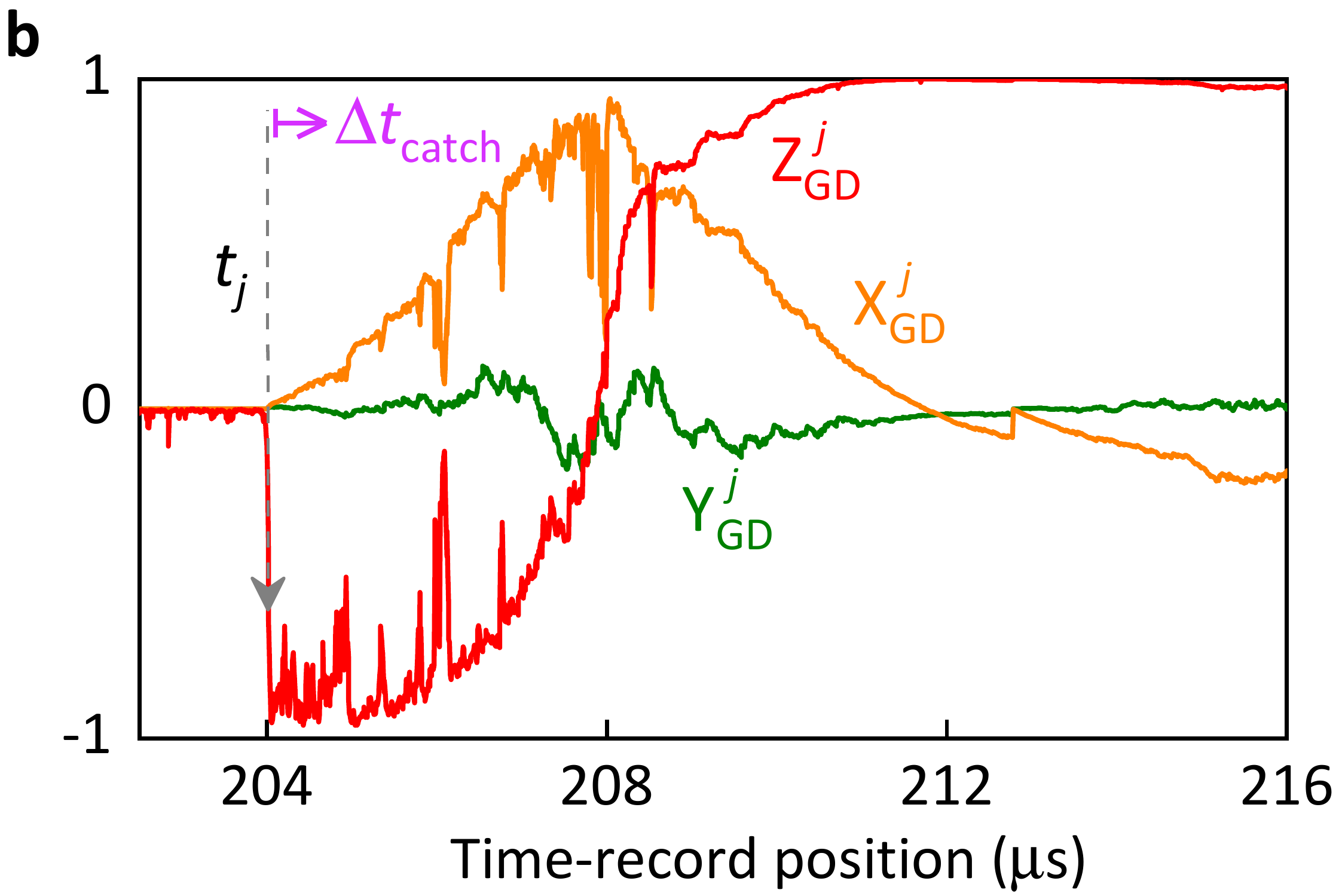}}
\caption{\label{fig:monte-carlo}
\textbf{Sampling of the Monte-Carlo simulation.} \textbf{a,} Simulated measurement quadrature $I_{\rm rec}$ and correlated trajectory computed from Eqs. (\ref{eqn:correlated_Z_GD}) and (\ref{eqn:correlated_X_GD&Y_GD}). Three sample intervals are shown. The earliest corresponds to leakage from the GBD-manifold, where a jump from $|{\rm G}\rangle$ to $|{\rm F}\rangle$ is followed by a jump from $|{\rm F}\rangle$ to $|{\rm D}\rangle$. The second and third sample intervals correspond to direct transitions from $|{\rm G}\rangle$ to $|{\rm D}\rangle$, which are continuously monitored and the object of the experiment. \textbf{b,} Expanded view of the shaded region of the second sample interval in panel (a). The evolution is continuous but not smooth, due to backaction noise from the continuously monitored readout. This feature is in sharp contrast to the perfect ``no-click'' readout upon which the simple theory of Sec.~\ref{sec:fluorescence_monitored_by_photon_counts} is based.
}
\vskip2.0cm

\hskip-1.75cm\includegraphics[width=9cm]{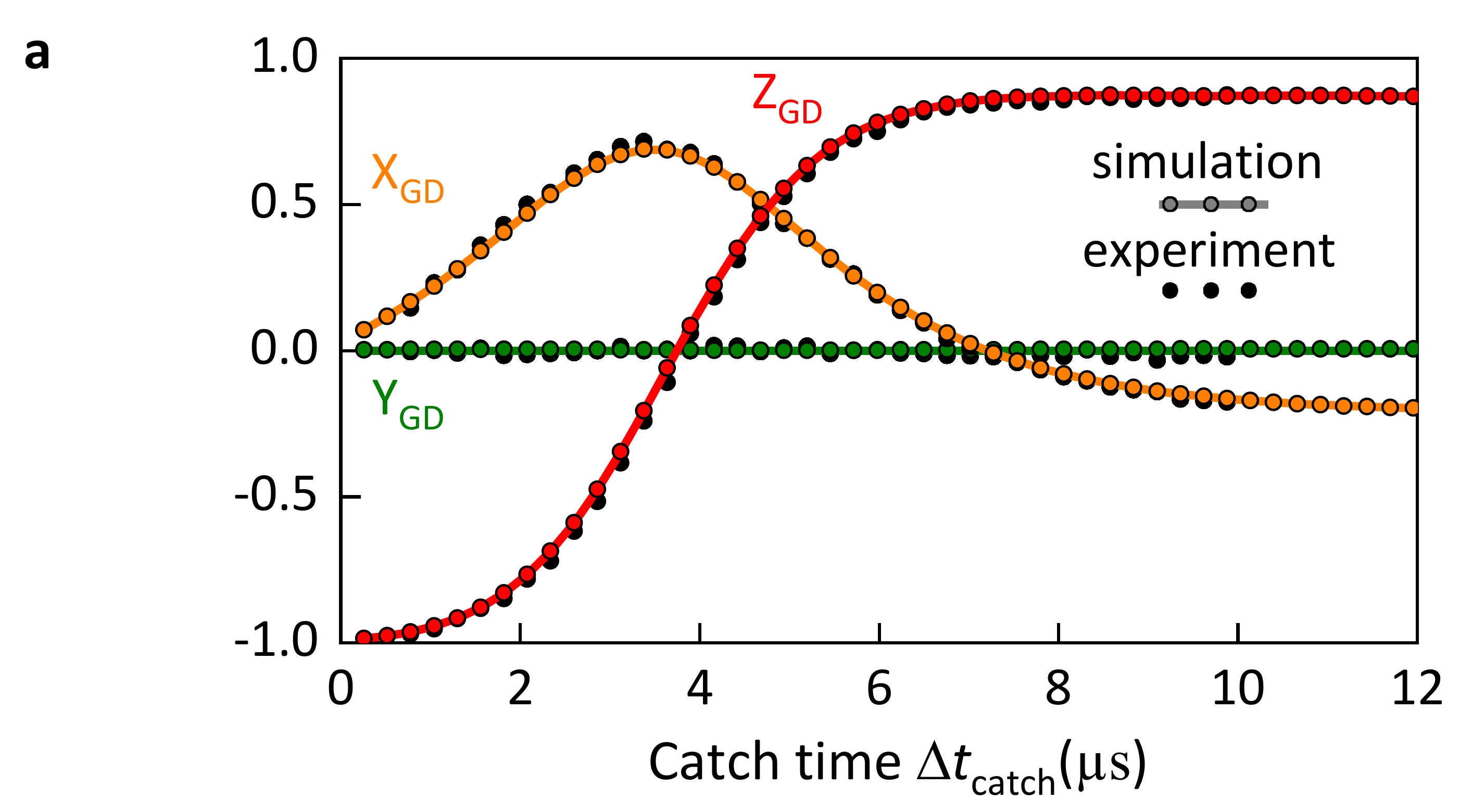}\hskip1cm
\vskip0.5cm
\hskip-1.75cm\includegraphics[width=9cm]{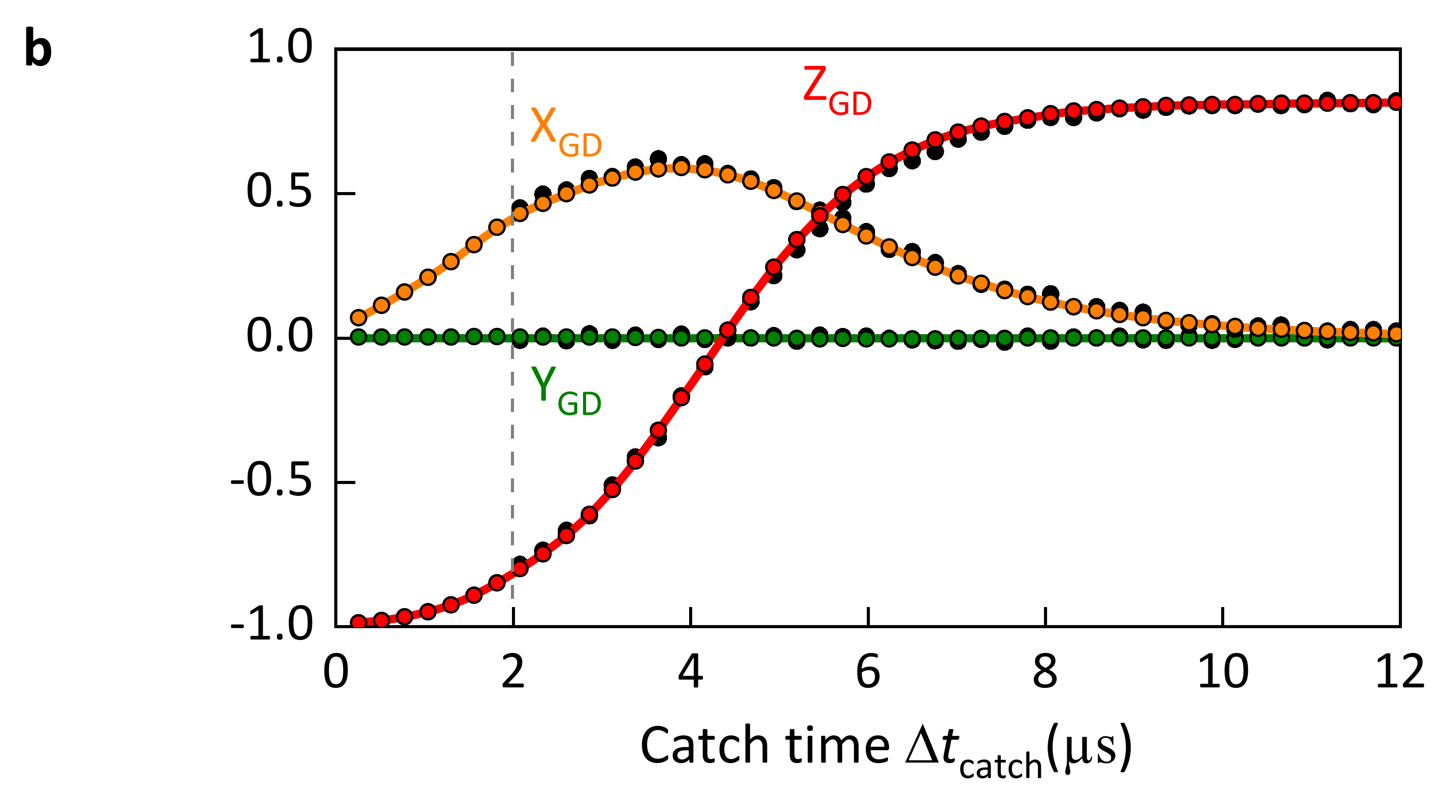}
\caption{\label{fig:simulation_vs_experiment}
\textbf{Comparison between simulation and experiment.} 
\textbf{a,}
 Simulated data set obtained with Rabi drive $\Omega_{\rm DG}$ turned on for the entire $\Delta t_{\rm catch}$; parameters taken from Table~\ref{table:table2} and leakage from the GBD-manifold included with $(\gamma_{\rm FG},\gamma_{\rm FD})/2\pi=0.38\mkern2mu{\rm kHz}$ and $(\gamma_{\rm GF},\gamma_{\rm DF})/2\pi=11.24\mkern2mu{\rm kHz}$. \textbf{b,} 
Simulated data set obtained with Rabi drive $\Omega_{\rm DG}$ turned off at time $\Delta t_{\rm on}=2\mkern2mu\mu{\rm s}$; parameters taken from Table \ref{table:table2} and leakage from the GBD-manifold included with $\gamma_{\rm FG}/2\pi=0.217\mkern2mu{\rm kHz}$, $\gamma_{\rm FD}/2\pi=4.34\mkern2mu{\rm kHz}$, $\gamma_{\rm GF}/2\pi=11.08\mkern2mu{\rm kHz}$, and $\gamma_{\rm DF}/2\pi=15.88\mkern2mu{\rm kHz}$. When leakage from the GBD-manifold is omitted, the ${\rm Z}_{\rm GD}$ curve rises more sharply and settles to a value that is  10\% (20\%) higher in panel (a) (panel (b)).
}
\end{centering}
\end{figure}

\textit{Ensemble average.} Simulated data sets are computed as an ensemble average by sampling an ongoing Monte Carlo simulation, numerically implementing the model outlined in Eqs.~(\ref{eqn:SSE_continuous})--(\ref{eqn:jumps_to_F}). Quadratures $I_{\rm rec}$ and $Q_{\rm rec}$ are computed from Eqs.~(\ref{eqn:simulated_I_int}) and (\ref{eqn:simulated_Q_int}), digitized with integration time $T_{\rm int}=260\mkern2mu{\rm ns}$, and then, as in the experiment, a hysteric filter is used to locate ``click'' events ($\Delta t_{\rm catch}=0$) corresponding to an inferred change of state from $|{\rm B}\rangle$ to not-$|{\rm B}\rangle$. During the subsequent sampling interval ($\Delta t_{\rm catch}\geq0$), the four quantities
\begin{equation}
\big({\rm Z}_{\rm GD}^j,{\rm X}_{\rm GD}^j,{\rm Y}_{\rm GD}^j,{\rm P}_{\rm BB}^j\big)(\Delta t_{\rm catch})=\big({\rm Z}_{\rm GD}^{\rm rec},{\rm X}_{\rm GD}^{\rm rec},{\rm Y}_{\rm GD}^{\rm rec},{\rm P}_{\rm BB}^{\rm rec}\big)(t_j+\Delta t_{\rm catch}),
\end{equation}
with $t_j$ is the click time and
\begin{eqnarray}
{\rm Z}_{\rm GD}^{\rm rec}(t)&=&\frac{\langle{\rm D}|\psi(t)\rangle\langle\psi(t)|{\rm D}\rangle-\langle{\rm G}|\psi(t)\rangle\langle\psi(t)|{\rm G}\rangle}{\langle\psi(t)|\psi(t)\rangle},\label{eqn:correlated_Z_GD}\\
\noalign{\vskip4pt}
{\rm X}_{\rm GD}^{\rm rec}(t)+i{\rm Y}_{\rm GD}^{\rm rec}(t)&=&2\frac{\langle{\rm D}|\psi(t)\rangle\langle\psi(t)|{\rm G}\rangle}{\langle\psi(t)|\psi(t)\rangle},\label{eqn:correlated_X_GD&Y_GD}\\
\noalign{\vskip4pt}
{\rm P}_{\rm BB}^{\rm rec}(t)&=&\frac{\langle{\rm B}|\psi(t)\rangle\langle\psi(t)|{\rm B}\rangle}{\langle\psi(t)|\psi(t)\rangle},
\end{eqnarray}
are computed, and running sums of each are updated. The sample terminates when the measurement record indicates a change of state from not-$|{\rm B}\rangle$ back to $|{\rm B}\rangle$. Finally, for comparison with the experiment, Bloch vector components are recovered from the average over sample intervals via the formula
\begin{equation}
\big({\rm Z}_{\rm GD},{\rm X}_{\rm GD},{\rm Y}_{\rm GD}\big)(\Delta t_{\rm catch})=\frac{\sum_j^{N(\Delta t_{\rm catch})}\big({\rm Z}_{\rm GD}^j,{\rm X}_{\rm GD}^j,{\rm Y}_{\rm GD}^j\big)(\Delta t_{\rm catch})}{N(\Delta t_{\rm catch})-\sum_j^{N(\Delta t_{\rm catch})}P_{\rm BB}^j(\Delta t_{\rm catch})},
\label{eqn:ensemble_average}
\end{equation}
where $N(\Delta t_{\rm catch})$ is the number of sample intervals that extend up to, or beyond, the time $\Delta t_{\rm catch}$. The simulation and sampling procedure is illustrated in Fig.~\ref{fig:monte-carlo}, and a comparison between the experiment and the simulation is provided in Fig.~\ref{fig:simulation_vs_experiment}.

The simulated and measured Bloch vector components are fit with expressions motivated by Eqs.~\eqref{eqn:Z_approx}-\eqref{eqn:Y_approx} and~\eqref{eq:XYZ_GD-inf}, modified to account for the effect of non-idealities in the experiment, 
\begin{align}
\mathrm{Z}_{\text{GD}}(\Delta t_{\operatorname{catch}}) & =\ensuremath{a+b\tanh(\Delta t_{\operatorname{catch}}/\tau+c)},\\
\ensuremath{\mathrm{X}_{\text{GD}}(\Delta t_{\operatorname{catch}})} & =a'+b'\operatorname{sech}(\Delta t_{\operatorname{catch}}/\tau'+c')\,,\\
\mathrm{Y}_{\text{GD}}(\Delta t_{\operatorname{catch}}) & =0\,.
\end{align}
The fit parameters  ($a,a',b,b',c,c',\tau,\tau'$) for the simulated and experimental data shown in Fig.~\ref{fig:simulation_vs_experiment} are compared in Table~\ref{tab:Comparison-of-parameters}. 
As imposed by Eq.~\eqref{eq:XYZ_GD-inf}, in the absence of $\Omega_{\rm DG}$ (turned off at time $\Delta t_{\rm on}=2\mkern2mu\mu{\rm s}$) $a'$, the  offset of $\mathrm{X}_{\text{GD}}$,  is strictly enforced to be zero.
The extracted simulation and experiment parameters are found to agree at the percent level.

\begin{table}[t]
\begin{tabular}{ r  c  l | r  c  l | r  c  l }
$\vphantom{\vbox{\vskip12pt}}\mkern40mu$&\textbf{Readout cavity}&$\mkern40mu$&$\mkern44mu$&\textbf{BG transition}&$\mkern44mu$&$\mkern44mu$&\textbf{DG transition}&$\mkern44mu$\\
\hline
\hline
\end{tabular}
\vskip12pt
\begin{tabular}{ r  c  l | r  c  l | r  c  l }
\multicolumn{9}{c}{$\vphantom{\vbox{\vskip10pt}}$\textbf{Non-linear parameters}\hfill}\\
\hline
&$\hphantom{\hbox{\hskip4.2cm}}\mkern-16mu$&&$\vphantom{\vbox{\vskip12pt}}\mkern32mu\chi_{\rm B}/2\pi$&$=$&$-5.08\mkern3mu{\rm MHz}\mkern32mu$&$\vphantom{\vbox{\vskip14pt}}\mkern30mu\chi_{\rm D}/2\pi$&$=$&$-0.33\mkern3mu{\rm MHz}\mkern30mu$\\
\end{tabular}
\vskip12pt
\begin{tabular}{ r  c  l | r  c  l | r  c  l }
\multicolumn{9}{c}{$\vphantom{\vbox{\vskip10pt}}$\textbf{Coherence related parameters}\hfill}\\
\hline
$\vphantom{\vbox{\vskip12pt}}\mkern46mu\kappa/2\pi$&$=$&$3.62\mkern3mu{\rm MHz}\mkern46mu$&$\mkern61muT_{1}^{\rm B}$&$=$&$15\mkern3mu\mu{\rm s}\mkern81mu$&$\mkern48muT_{1}^{\rm D}$&$=$&$105\mkern3mu\mu{\rm s}\mkern74mu$\\
$\vphantom{\vbox{\vskip12pt}}\eta$&$=$&$0.33$&$T_{2}^{\rm B}$&$=$&$18\mkern3mu\mu{\rm s}$&$T_{2}^{\rm D}$&$=$&$120\mkern3mu\mu{\rm s}$\\
$\vphantom{\vbox{\vskip12pt}}T_{\rm int}$&$=$&$260.0\mkern3mu{\rm ns}$ & &&&&&\\
$\vphantom{\vbox{\vskip12pt}}n_{\rm th}^{\rm C}$&$=$&$0$&$n_{\rm th}^{\rm B}$&$=$&$0.01$&$n_{\rm th}^{\rm D}$&$=$&$0.05$\\
\end{tabular}
\vskip12pt
\begin{tabular}{ r  c  l | r  c  l | r  c  l }
\multicolumn{9}{c}{$\vphantom{\vbox{\vskip10pt}}$\textbf{Drive amplitude and detuning parameters}\hfill}\\
\hline
$\vphantom{\vbox{\vskip12pt}}\mkern72mu\bar{n}$&$=$&$5.0\mkern40mu\mkern56mu$&$\Omega_{{\rm B}0}/2\pi$&$=$&$1.2\mkern3mu{\rm MHz}$&$\Omega_{\rm DG}/2\pi$&$=$&$21.6\mkern3mu{\rm kHz}$\\
$\vphantom{\vbox{\vskip12pt}}$&&&$\mkern20mu\Omega_{{\rm B}1}/2\pi$&$=$&$600\mkern3mu{\rm kHz}\mkern30mu$&&&\\
$\vphantom{\vbox{\vskip12pt}}\Delta_{\rm R}$&$=$&$\chi_{\rm B}$&$\mkern22mu\Delta_{{\rm B}1}/2\pi$&$=$&$-30.0\mkern3mu{\rm MHz}\mkern31mu$&$\mkern19mu\Delta_{\rm DG}/2\pi$&$=$&$-274.5\mkern3mu{\rm kHz}\mkern30mu$\\
\end{tabular}
\caption{\textbf{Compilation of the simulation parameters.}}
\label{table:table2}
\end{table}

\subsection{Error budget}
\label{sec:Error-budget}

\subsubsection{Imperfections}
Various imperfections are expected to reduce the maximum coherence recovered in the measurement of ${\rm X}_{\rm GD}(\Delta t_{\rm catch})$. They include: \begin{enumerate}[(i)]
\item Readout errors when inferring $|{\rm B}\rangle$ to not-$|{\rm B}\rangle$ transitions and the reverse. Such errors affect the assignment of $\Delta t_{\rm catch}$, which can be either too short or too long to correlate correctly with the true state of the system.
\item Leaks from the GBD-manifold to higher excited states. These errors mimic a $|{\rm B}\rangle$ to not-$|{\rm B}\rangle$ transition, as in the first sample interval of Fig.~\ref{fig:monte-carlo}, yet the anticipated coherent evolution within the GBD-manifold does not occur.
\item Thermal jumps from $|{\rm G}\rangle$ to $|{\rm D}\rangle$. Such incoherent transitions contribute in a similar way to ${\rm Z}_{\rm GD}(\Delta t_{\rm catch})$, while making no contribution to the measured coherence.
\item Direct dephasing of the DG-coherence.
\item Partial distinguishability of $|{\rm G}\rangle$ and $|{\rm D}\rangle$. The readout cavity is not entirely empty of photons when the state is not-$|{\rm B}\rangle$, in which case the cross-Kerr interaction $\chi_{\rm D}|{\rm D}\rangle\langle{\rm D}|\hat c^\dagger\hat c$ shifts the $\Omega_{\rm DG}$ Rabi drive from resonance; hence, backaction noise is transferred from the photon number to ${\rm X_{\rm GD}}(\Delta t_{\rm catch})$.
\end{enumerate}

\subsubsection{Budget for lost coherence}
\label{sec:Budget-coherence}
\begin{figure}[!ht]
\begin{centering}
\includegraphics[height=4.8cm]{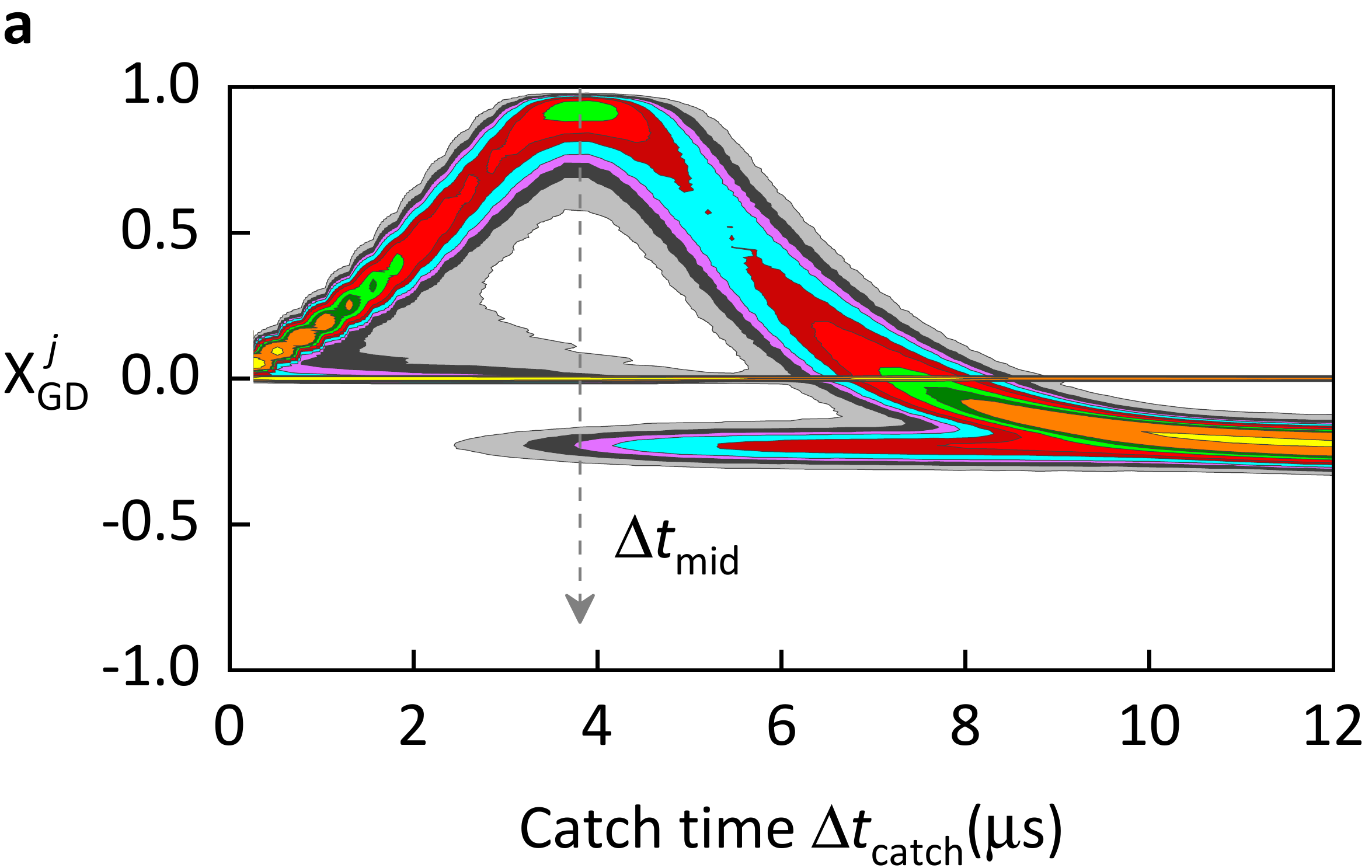}\hskip0.25cm\hbox{\includegraphics[height=4.8cm]{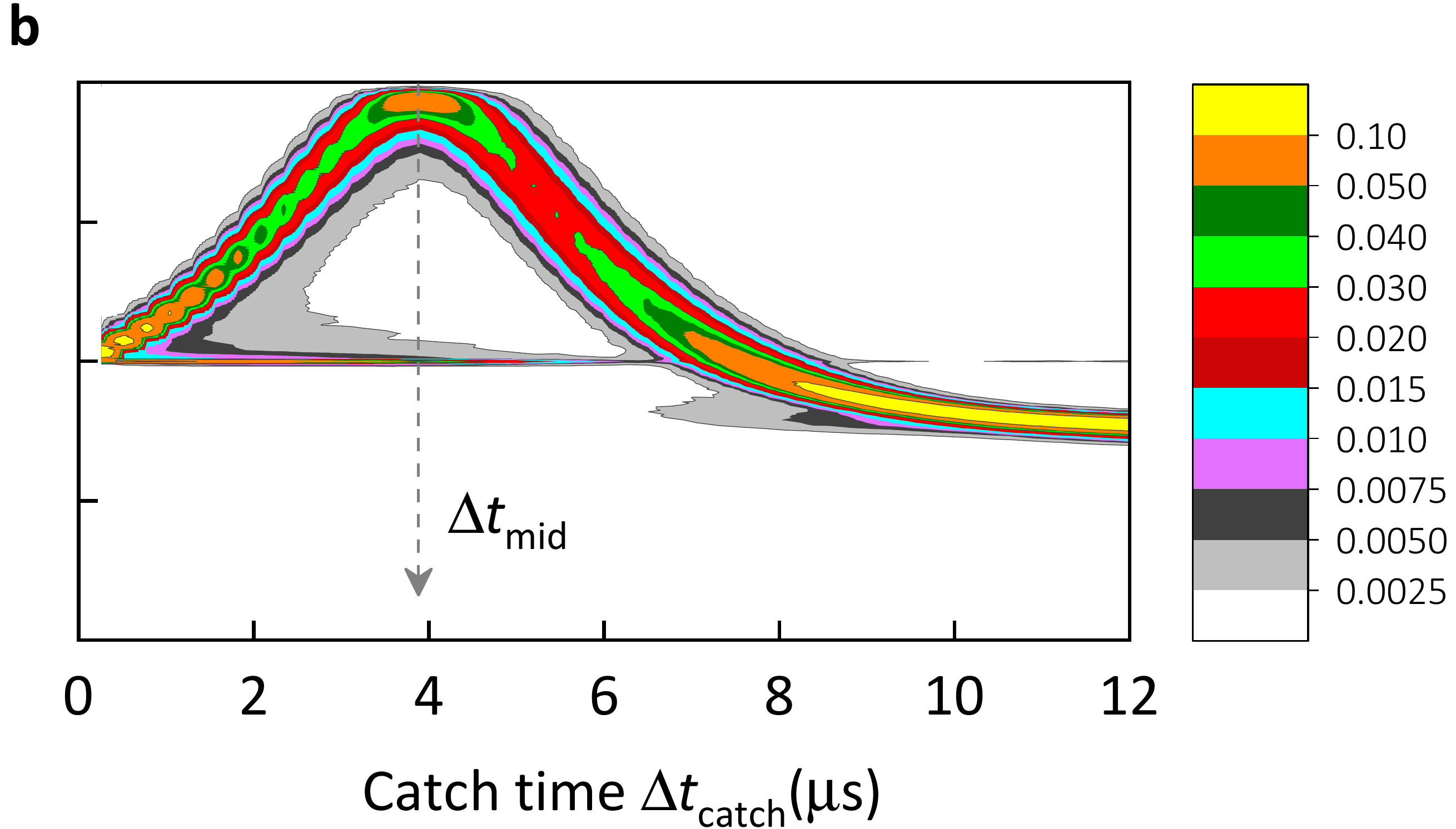}}
\caption{\label{fig:coherence_loss}
\textbf{Coherence loss through sample to sample fluctuations.} \textbf{a,} Contour plot of the distribution of ${\rm X}_{\rm GD}^j(\Delta t_{\rm catch})$ samples corresponding to the simulated data set displayed in panel (a) of Fig.~\ref{fig:simulation_vs_experiment}. \textbf{b,} Same as panel (a) but with transitions between $|{\rm B}\rangle$ and not-$|{\rm B}\rangle$ identified in the ket $|\psi\rangle$ rather than from the simulated measurement record, and with changed parameters: $(\gamma_{\rm FG},\gamma_{\rm FD},\gamma_{\rm GF},\gamma_{\rm DF})/2\pi=0$, $n_{\rm th}^{\rm B}=n_{\rm th}^{\rm D}=0$, $T_2^{\rm D}=2T_1^{\rm D}$, and $\chi_{\rm D}/2\pi=0$.
}
\end{centering}
\end{figure}

The maximum coherence reported in the experiment is $0.71\pm0.005$. In the simulation it is a little lower at 0.69. By removing the imperfections from the simulation, one by one, we can assign a fraction of the total coherence loss to each. Readout errors are eliminated by identifying transitions between $|{\rm B}\rangle$ and not-$|{\rm B}\rangle$ in the ket $|\psi\rangle$ rather than from the simulated measurement record; all other imperfections are turned off by setting some parameter to zero. The largest coherence loss comes from readout errors, whose elimination raises the ${\rm X}_{\rm GD}(\Delta t_{\rm catch})$ maximum by 0.09. The next largest comes from leakage to higher excited states, which raises the maximum by a further 0.06. Setting $\chi_{\rm D}$ to zero adds a further 0.04, and thermal transitions and pure dephasing together add 0.02. Figure \ref{fig:coherence_loss} illustrates the change in the distribution of ${\rm X}_{\rm GD}^j(\Delta t_{\rm catch})$ samples underlying the recovery of coherence. The removal of the finger pointing to the left in panel (a) is mainly brought about by the elimination of readout errors, while the reduced line of zero coherence marks the elimination of leakage to higher excited states. Aside from these two largest changes, there is also a sharpening of the distribution, at a given $\Delta t_{\rm catch}$, when moving from panel (a) to panel (b). Having addressed the five listed imperfections, a further 10\% loss remains unaccounted for, i.e., the distribution of panel (b) is not a line passing through ${\rm X}_{\rm GD}^j(\Delta t_{\rm mid})=1$. The final 10\% is explained by the heterodyne detection backaction noise,  a function of the drive and measurement parameters, displayed in panel (b) of Fig.~\ref{fig:monte-carlo}.

\begin{table}
\begin{centering}
\begin{tabular}{>{\centering}p{0.5\textwidth}>{\centering}p{0.5\textwidth}}
     \textbf{(a)} In presence of $\Omega_{\mathrm{DG}}$  \\
 & \textbf{(b)} In absence of $\Omega_{\mathrm{DG}}$  \\
\tabularnewline
\centering{}\renewcommand*\arraystretch{1.2}
\begin{tabular}{cc|crcllrrlcr}
Parameter & \multicolumn{1}{c}{} &  & \multicolumn{3}{c}{Experiment} &  & \multicolumn{3}{c}{Simulation} &  & Error\tabularnewline
\hline 
\rule{0pt}{3ex} $a$ &  &  & -0.07 & $\pm$ & 0.005 &  & -0.07 & $\pm$ & 0.005 &  & 0.5\%\tabularnewline
$a'$ &  &  & -0.21 & $\pm$ & 0.005 &  & -0.22 & $\pm$ & 0.005 &  & 2\%\tabularnewline
$b$ &  &  & 0.94 & $\pm$ & 0.005 &  & 0.95 & $\pm$ & 0.005 &  & 1\%\tabularnewline
$b'$ &  &  & 0.93 & $\pm$ & 0.005 &  & 0.91 & $\pm$ & 0.005 &  & 2\%\tabularnewline
$c$ &  &  & -2.32 & $\pm$ & 0.03 &  & -2.27 & $\pm$ & 0.03 &  & 2\%\tabularnewline
$c'$ &  &  & -2.04 & $\pm$ & 0.03 &  & -2.05 & $\pm$ & 0.03 &  & 0.5\%\tabularnewline
$\tau$ &  &  & 1.64 & $\pm$ & 0.01 &  & 1.65 & $\pm$ & 0.01 &  & 0.5\%\tabularnewline
$\tau'$ &  &  & 1.74 & $\pm$ & 0.01 &  & 1.76 & $\pm$ & 0.01 &  & 1\%\tabularnewline
\end{tabular} & \centering{}\renewcommand*\arraystretch{1.2}
\begin{tabular}{cc|crcllrrlcr}
Parameter & \multicolumn{1}{c}{} &  & \multicolumn{3}{c}{Experiment} &  & \multicolumn{3}{c}{Simulation} &  & Error\tabularnewline
\hline 
\rule{0pt}{3ex} $a$ &  &  & -0.11 & $\pm$ & 0.005 &  & -0.10 & $\pm$ & 0.005 &  & 8\%\tabularnewline
$a'$ &  &  & 0 & $\pm$ & 0 &  & 0 & $\pm$ & 0 &  & 0\%\tabularnewline
$b$ &  &  & 0.92 & $\pm$ & 0.008 &  & 0.91 & $\pm$ & 0.008 &  & 1\%\tabularnewline
$b'$ &  &  & 0.61 & $\pm$ & 0.005 &  & 0.60 & $\pm$ & 0.005 &  & 2\%\tabularnewline
$c$ &  &  & -1.96 & $\pm$ & 0.05 &  & -2.10 & $\pm$ & 0.05 &  & 7\%\tabularnewline
$c'$ &  &  & -1.97 & $\pm$ & 0.05 &  & -2.05 & $\pm$ & 0.05 &  & 4\%\tabularnewline
$\tau$ &  &  & 2.17 & $\pm$ & 0.05 &  & 2.03 & $\pm$ & 0.05 &  & 6\%\tabularnewline
$\tau'$ &  &  & 1.98 & $\pm$ & 0.05 &  & 1.92 & $\pm$ & 0.05 &  & 3\%\tabularnewline
\end{tabular}\tabularnewline
\end{tabular}
\par\end{centering}
\caption{\label{tab:Comparison-of-parameters} \textbf{Comparison between 
parameters extracted from the simulation and those from the experiment}. 
\textbf{a,}
Parameters obtained from fits of the simulated and measured data for the catch protocol
in the presence of the Rabi drive $\Omega_{{\rm DG}}$ throughout the
entire duration of the quantum jump, data shown in Fig.~\ref{fig:simulation_vs_experiment}a. 
\textbf{b,} Parameters obtained
from fits of the simulated and measured data
for the catch protocol in the absence of the $\Omega_{{\rm DG}}$
during the flight of the quantum jump for $\Delta t_{\mathrm{on}}=2\mathrm{\ \mu s}$, data shown in  Fig.~\ref{fig:simulation_vs_experiment}b.
}
\end{table}

\subsection{Signal-to-noise ratio (SNR) and de-excitation measurement efficiency\label{subsec:Signal-to-noise-ratio-(SNR)}}

As discussed in the Methods section, the catch protocol hinges on
the efficient detection of de-excitations from $|{\rm B}\rangle$
to $|{\rm G}\rangle$. In atomic physics, de-excitations are typically
monitored by a \emph{direct} detection method, employing a photodetector.
Alternatively, de-excitations can be monitored by an\emph{ indirect}
method, as done in our experiment. In this subsection, we discuss
the efficiency of both methods. For the indirect method, using simple
analytics, we estimate the \emph{total} efficiency of time-continuous,
uninterrupted monitoring of de-excitations from $|{\rm B}\rangle$
to $|{\rm G}\rangle$ to be $\eta_{\mathrm{eff,clk}}=0.90\pm0.01$
for the parameters of our experiment, with integration time $T_{\mathrm{int}}=0.26\,\mathrm{\mu s}$.
The simple analysis of this section complements the numerical one
of the previous section, Sec.~\ref{sec:Budget-coherence}.

\emph{Direct monitoring method in atomic physics.} The direct method
monitors for a $|{\rm B}\rangle$ de-excitation by collecting and
absorbing the photon radiated in the de-excitation. The \emph{total}
measurement efficiency of this method is limited by i) collection
efficiency \textemdash{} the fraction of emitted photons collected
by the detector in its own input spatial modes (for instance, as determined
by the solid angle) \textemdash{} typically falls in the range 0.1
- 50\%,\cite{Volz2011} ii) the efficiency of detecting the absorption
of a single photon, which falls in the range 1 - 90\%,\cite{Eisaman2011}
and iii) non-idealities of the photodetector apparatus, including
its dead time, dark counts, jitter, etc.\cite{Eisaman2011} The combination
of these inefficiencies presents an almost insurmountable challenge
in experimental atomic physics for realizing continuous, time-resolved
detection of nearly every single photon emitted by the three-level
atom, required to faithfully catch the jump. 

\emph{Direct monitoring method with superconducting circuits.} While
technologically very different, the direct monitoring method with
superconducting circuits is conceptually similar to atomic method
but can readily achieve high collection efficiencies.\cite{Katz2008,Vijay2011,Riste2012-qubit-measure-reset,Vijay2012,Hatridge2013,Murch2013a,deLange2014,Roch2014,Weber2014,Campagne-Ibarcq2014,Macklin2015,Campagne2016-Fluorescence,Campagne-Ibarcq2016,Hacohen-Gourgy2016-non-comm,Naghiloo2016,White2016,Ficheux2017,Naghiloo2017-thermo,Tan2017,Hacohen-Gourgy2018,Heinsoo2018,Bultink2018}
 However, the energy of the emitted microwave photon is exceedingly
small \textemdash{} $23\text{ }\mathrm{\mu eV}$, about a part per
100,000 of the energy of a single optical photon \textemdash{} which
essentially forbids the direct detection of the photon with near-unit
efficiency. This is because the propagating photon is unavoidably
subjected to significant loss, added spurious noise, amplifier non-idealities,
etc. In our experiment, these imperfections reduce the full measurement/amplification
chain efficiency from its ideal value\cite{Hatridge2013,Macklin2015,Bultink2018}
of 1 to a modest $\eta=0.33\pm0.03$, corresponding to the direct
detection of approximately only one out of every three single photons
\textemdash{} insufficient for the catch protocol.

\subsubsection{Indirect monitoring method with superconducting circuits}

Alternatively, the indirect monitoring method couples the atom to
an ancillary degree of freedom, which is itself monitored in place
of the atom. In our experiment, the atom is strongly, dispersively
coupled to the ancillary readout cavity. The cavity scatters a probe
tone, whose phase shift constitutes the readout signal, as discussed
in the Methods section. Since the probe tone can carry itself many
photons, this scheme increases the signal-to-noise ratio ($\mathrm{SNR}$)
and, hence, the total efficiency ($\eta_{\mathrm{eff,clk}}$) of detecting
a $|{\rm B}\rangle$ de-excitation. Note that the efficiency $\eta_{\mathrm{eff,clk}}$
should not be confused with the efficiency of a photodetector or the
efficiency $\eta$ of the measurement/amplification chain, since $\eta_{\mathrm{eff,clk}}$
includes the effect of all readout imperfections and non-idealities,
state discrimination and assignment errors, etc. see below. In the
remainder of this section, we estimate the SNR and efficiency $\eta_{\mathrm{eff,clk}}$
of the experiment.

\emph{SNR of the indirect (dispersive) method.}  The output of the
measurement and amplification chain monitoring the readout cavity
is proportional to the complex heterodyne measurement record $\zeta\left(t\right)$,
which obeys the It\^{o} stochastic differential equation, see Eq.~\eqref{eq:heterodyne-current},\footnote{Since the bandwidth of the measurement chain, $\kappa_{\mathrm{filter}}$,
is significantly larger than that, $\kappa$, of the readout cavity,
$\kappa_{\mathrm{filter}}\gg\kappa$, we can neglect the effect of
$\kappa_{\mathrm{filter}}$ for simplicity of discussion, see Eqs.~\eqref{eq:dI_rec}
and~\eqref{eq:dQ_rec}. } 
\begin{equation}
\mathrm{d}\zeta\left(t\right)=\sqrt{\eta\kappa}\frac{\langle\psi\left(t\right)|\hat{a}|\psi\left(t\right)\rangle}{\langle\psi\left(t\right)|\psi\left(t\right)\rangle}\mathrm{d}t+\mathrm{d}Z\left(t\right),\label{eq:heterodyne-current2}
\end{equation}
where $\hat{a}$ is the cavity amplitude operator in the Schr\"{o}dinger
picture, $\eta$ is the total measurement efficiency of the amplification
chain \textemdash{} again, not to be confused with the de-excitation
measurement efficiency, $\eta_{\mathrm{eff,clk}}$ \textemdash{} and
$\mathrm{d}Z$ is the complex Wiener process increment, defined below
Eq.~\eqref{eq:heterodyne-current2}. A somewhat counterintuitive
property of Eq.~\eqref{eq:heterodyne-current2} is that  the heterodyne
record increment $\mathrm{d}\zeta\left(t\right)$ is stochastic and
noisy even when $\eta=1$, the case of ideal measurement in which
no signal is lost \textemdash{} the stochastic term, $\mathrm{d}Z$,
represents pure quantum vacuum fluctuations, which are inherent in
the case of heterodyne detection.\cite{Carmichael1993,Plenio1998,wiseman2010book}
Due to the unavoidable presence of these fluctuations, only an infinitesimal
amount of information about the system can be extracted from $\mathrm{d}\zeta$
at an instant of time. Finite amount of information is extracted by
integrating $\mathrm{d}\zeta$ for a finite duration $T_{\mathrm{int}}$,
\begin{equation}
s\equiv I_{\mathrm{rec}}+iQ_{\mathrm{rec}}\equiv\int_{0}^{T_{\mathrm{int}}}\mathrm{d}\zeta\left(t\right)\,,\label{eq:s=00003D}
\end{equation}
where $I_{\mathrm{rec}}$ and $Q_{\mathrm{rec}}$ are the in- and
out-of-phase quadrature components of one segment of the record. What
does $s$ correspond to? Its value depends on $\mathrm{d}\zeta$,
which depends on the state of the cavity, $|\psi\rangle$, which itself
depends on the occupation of $\ket{\mathrm{B}}$ \textemdash{} and
therefore $s$ contains the occupation of $\ket{\mathrm{B}}$. A de-excitation
of $\ket{\mathrm{B}}$ to $\ket{\mathrm{G}}$ can thus be detected
by monitoring $s$, whose value is different for the two states, since
the cavity is generally in the coherent state $\ket{\alpha_{\mathrm{B}}}$
or $\ket{\alpha_{\mathrm{G}}}$ when the atom is in $\ket{\mathrm{B}}$
or $\ket{\mathrm{G}}$, respectively. For the moment, assuming the
atom and cavity do not change states during the course of the measurement
duration $T_{\mathrm{int}}$, the stochastic integral in Eq.~\eqref{eq:s=00003D}
explicitly evaluates to 
\begin{equation}
s_{\mathrm{B,G}}=\left\{ \sqrt{\eta\kappa}\mathrm{Re}\left[\alpha_{\mathrm{B,G}}\right]T_{\mathrm{int}}+\frac{1}{\sqrt{2}}W_{I}\left(T_{\mathrm{int}}\right)\right\} +i\left\{ -\sqrt{\eta\kappa}\mathrm{Im}\left[\alpha_{\mathrm{B,G}}\right]T_{\mathrm{int}}+\frac{1}{\sqrt{2}}W_{Q}\left(T_{\mathrm{int}}\right)\right\} \,,\label{eq:s=00003DExplicit}
\end{equation}
where $W_{I,Q}$ denote independent Wiener processes, obeying the
conventional rules, $\mathrm{E}\left[W\left(t\right)\right]=0$ and
$\mathrm{Var}\left[W\left(t\right)\right]=t^{2}$. Equation~\eqref{eq:s=00003DExplicit}
shows that the distribution of the stochastic variable $s$ is a Gaussian
blob in the IQ plane centered at $\bar{s}_{\mathrm{B,G}}\equiv\operatorname{E}\left[s_{\mathrm{B,G}}\right]=\sqrt{\eta\gamma}T_{\mathrm{int}}\alpha_{\mathrm{B,G}}$
with width determined by the variance $\sigma_{\mathrm{B,G}}^{2}\equiv\operatorname{Var}\left[s_{\mathrm{B,G}}\right]=\frac{1}{2}T_{\mathrm{int}}$.
We can thus define the SNR of the experiment by comparing the distance
between the two pointer distributions to their width, 
\begin{equation}
\mathrm{SNR}\equiv\left|\frac{\bar{s}_{\mathrm{B}}-\bar{s}_{\mathrm{G}}}{\sigma_{\mathrm{B}}+\sigma_{\mathrm{G}}}\right|^{2}\,,\label{eq:SNR-defn}
\end{equation}
where the B (resp., G) subscript denotes signals conditioned on the
atom being in $\ket{\mathrm{B}}$ (resp., $\ket{\mathrm{G}}$). In
terms of $\ket{\alpha_{\mathrm{B}}}$ and $\ket{\alpha_{\mathrm{G}}}$,
\begin{equation}
\mathrm{SNR}=\frac{1}{2}\eta\kappa T_{\mathrm{int}}\left|\alpha_{\mathrm{B}}-\alpha_{\mathrm{G}}\right|^{2}\,,
\end{equation}
which can be expressed in terms of the parameters of the experiment,
summarized in Table~\ref{tab:system-params},
\begin{equation}
\mathrm{SNR}=\frac{1}{2}\eta\kappa T_{\mathrm{int}}\left[\cos\left(\arctan\left(\frac{\kappa}{2\chi_{\mathrm{BG}}}\right)\right)\right]^{2}\bar{n}\,,\label{eq:SNR-expression}
\end{equation}
Holding other parameters fixed, according to Eq.~\eqref{eq:SNR-expression},
the SNR can be increased arbitrarily by increasing $\bar{n}$, which
can be readily done by increasing the amplitude of the cavity probe
tone. A higher SNR for $s$ corresponds to a higher SNR for measuring
an atom de-excitation, since $s$ is a proxy of the $\ket{\mathrm{B}}$
population. Thus, the indirect cavity monitoring can overcome the
typical degradation in SNR imposed by the inefficiencies and non-idealities
of the measurement chain, $\eta$. In practice, the SNR increase
with $\bar{n}$ is bounded from above, since with sufficiently high
$\bar{n}$ spurious non-linear effects become significant\cite{Boissonneault2008,Boissonneault2009-Photon-induced-relax,Minev2013,Sank2016-T1vsNbar,Khezri2016,Bultink2016,Khezri2017,Walter2017,Lescanne2018,Verney2018,Serniak2018}.
The cavity and non-linear coupling to the atom serve in effect as
a rudimentary embedded pre-amplifier at the site of the atom, which
transduces with amplification the de-excitation signal before its
SNR is degraded during propagation and further processing.

\emph{Discrimination efficiency of the indirect method.} While the
SNR provides a basic characterization of the measurement, it is useful
to convert it to a number between 0 and 1, which is called the discrimination
efficiency, $\eta_{\mathrm{disc}}$. It quantifies the degree to which
the two Gaussian distributions of $s$ are distinguishable,\cite{Gambetta2007-ProtocolsMsr}
\begin{equation}
\eta_{\mathrm{disc}}=\frac{1}{2}\operatorname{erfc}\left[-\sqrt{\frac{\mathrm{SNR}}{2}}\right]\,,\label{eq:eta-msr}
\end{equation}
where $\operatorname{erfc}$ denotes the complementary error function.
Equation~\eqref{eq:eta-msr} shows that increasing the SNR by separating
the $s_{\mathrm{B}}$ and $s_{\mathrm{G}}$ distributions far beyond
their spread, $\sigma_{\mathrm{B/G}}$, provides only marginal gain
as $\eta_{\mathrm{disc}}$ saturates to 1. Next, we calculate the
SNR and $\eta_{\mathrm{disc}}$ for the parameters of the experiment
and discuss corrections due to readout non-idealities. 

\emph{A first comparison to the experiment. }A first estimate of the
SNR and $\eta_{\mathrm{disc}}$ of the experiment are provided by
Eqs.~\eqref{eq:SNR-expression} and~\eqref{eq:eta-msr}. Using the
parameters of the experiment, summarized in Table~\ref{tab:system-params},
from these two equations, we find $\mathrm{SNR}=4.3\pm0.6$ and $\eta_{\mathrm{disc}}=0.98\pm0.007$.
Using data from the experiment, in particular, a second long IQ record
trace, represented by a short segment in Fig.~2a, we find the SNR
of the jumps experiment, by fitting the histogram of the trace with
a bi-Gaussian distribution, to be $\mathrm{SNR}=3.8\pm0.4$, corresponding
to $\eta_{\mathrm{\mathrm{disc}}}=0.96\pm0.01$. The measured values
are slightly lower than the analytics predict due to readout imperfections
not included in the calculation so far, such as state transitions
during $T_{\mathrm{int}}$, cavity transient dynamics, additional
pointer-state distributions, etc.

\emph{Effective click detection efficiency. }The dominant next-order
error is due to atom state transitions during the measurement window,
$T_{\mathrm{int}}$, which contributes an assignment error of approximately
$1-\eta_{\mathrm{asg}}=1-\exp\left(T_{\mathrm{int}}/\tau_{\mathrm{B}}\right)=0.06\pm0.001$
to the detection of a $|\mathrm{B}\rangle$ de-excitation. Combining
$\eta_{\mathrm{disc}}$ with $\eta_{\mathrm{asg}}$, we obtain the
total efficiency for detecting $\ket{\mathrm{B}}$ de-excitations
$\eta_{\mathrm{eff,clk}}=\eta_{\mathrm{disc}}\eta_{\mathrm{asg}}=0.90\pm0.01$,
consistent with the total readout efficiency of $0.91$ that is independently
estimated using the trajectory numerics, see Sec.~\ref{sec:Budget-coherence}.

\nocite{apsrev41Control}
\bibliographystyle{apsrev4-1}
\bibliography{mycontrol,biblio}